\documentclass[11pt,draft]{IEEEtran.OLD}
\usepackage{times}
\usepackage{cite} 
\usepackage{amssymb} 
\usepackage{bm}
\usepackage{subfigure}

\newcommand{\Cplx}      {{{\mathsf I}\!\!\!{\mathrm C}}} 
\newcommand{\NCplx}     {{{\mathcal{CN}}}} 

\newcommand{\define}    {\stackrel{\triangle}{=}}  
\newcommand{\tr}      {{\mathrm{tr}}}    
\newcommand{\E}       {{\mathrm E}}      

\newcommand{\Zrb}     {{\mathbf 0}}      

\newcommand{\ab}{{\mathbf a}}  \newcommand{\Ab}{{\mathbf A}} 
\newcommand{\bb}{{\mathbf b}}  \newcommand{\Bb}{{\mathbf B}} 
  \newcommand{\Cb}{{\mathbf C}} 
\newcommand{\db}{{\mathbf d}}  \newcommand{\Db}{{\mathbf D}}

\newcommand{\hb}{{\mathbf h}}  \newcommand{\Hb}{{\mathbf H}} 
  \newcommand{\Ib}{{\mathbf I}} 
   
  \newcommand{\Kb}{{\mathbf K}}
  
\newcommand{\mb}{{\mathbf m}}  
  \newcommand{\Nb}{{\mathbf N}}
  
  \newcommand{\Pb}{{\mathbf P}}
\newcommand{\qb}{{\mathbf q}}  \newcommand{\Qb}{{\mathbf Q}}
  \newcommand{\Rb}{{\mathbf R}}
\newcommand{\Sb}{{\mathbf S}} 

\newcommand{\vb}{{\mathbf v}}  \newcommand{\Vb}{{\mathbf V}}
\newcommand{\wb}{{\mathbf w}}  
\newcommand{\xb}{{\mathbf x}}  \newcommand{\Xb}{{\mathbf X}}
\newcommand{\yb}{{\mathbf y}}  \newcommand{\Yb}{{\mathbf Y}}
\newcommand{\zb}{{\mathbf z}}  

\newcommand{\mub}         {{\bm \mu}}

            \newcommand{\Phib}     {{\bm \Phi}}

         \newcommand{\Ak} {{\bm {\mathcal A}}}
         
         \newcommand{\Ck} {{\bm {\mathcal C}}}
         \newcommand{\Dk} {{\bm {\mathcal D}}}

         \newcommand{\Pk} {{\bm {\mathcal P}}}

         \newcommand{\Sk} {{\bm {\mathcal S}}}

\newcommand{\ba}{\begin{array}}
\newcommand{\ea}{\end{array}}

\usepackage[final]{graphicx}
\newtheorem{definition}{Definition}

\newtheorem{theorem}{Theorem}
\newtheorem{lemma}{Lemma}

\newtheorem{corollary}{Corollary}

\begin{document}
 \title{\LARGE \bf {Optimal Constellations for the Low SNR Noncoherent MIMO Block Rayleigh Fading Channel}\thanks{This work was supported in part by
     NSF grants CCF-0434410 and CCF-0431170.  It was presented in part at the Allerton Conf. Commun., Cntrl. and Comput., Monticello, IL, Sept. 2006 and the Intnl. Symp. on Inform. Th., Nice, France, June 2007. The authors are with the ECE department of the University of Colorado at Boulder, CO, 80309-0425.
     This paper was submitted Mar. 9, 2007, revised Mar. 21, 2008 and accepted Aug. 15, 2008.}}
 \author{
  Shivratna G. Srinivasan and  Mahesh K.  Varanasi
 }
  \maketitle \thispagestyle{empty}  \markboth{IEEE
  Trans. Inform. Th.}{Srinivasan \ \& \ Varanasi}

 \begin{abstract}
 Reliable communication over the discrete-input/continuous-output noncoherent multiple-input multiple-output (MIMO) Rayleigh block fading channel is considered when the signal-to-noise ratio (SNR) per degree of freedom is low. Two key problems are posed and solved to obtain the optimum discrete input. In both problems, the average and peak power per space-time slot of the input constellation are constrained. In the first one, the peak power to average power ratio (PPAPR) of the input constellation is held fixed, while in the second problem, the peak power is fixed independently of the average power. In the first {\em PPAPR-constrained} problem, the mutual information, which grows as ${\rm O(SNR^2)}$, is maximized up to second order in SNR. In the second {\em peak-constrained} problem, where the mutual information behaves as ${\rm O(SNR)}$, the structure of  constellations that are optimal up to first order, or equivalently, that minimize energy/bit, are  explicitly characterized. Furthermore, among constellations
that are first-order optimal, those that maximize the mutual information up to second order, or equivalently, the wideband slope, are characterized. In both PPAPR-constrained and peak-constrained problems, the optimal constellations are obtained in closed-form as solutions to non-convex optimizations, and interestingly, they are found to be identical. Due to its special structure, the common solution is referred to as Space Time Orthogonal Rank one Modulation, or STORM. In both problems, it is seen that STORM provides a sharp characterization of the behavior of noncoherent MIMO capacity.
 

 \end{abstract}
 
 \section{Introduction}

In this paper, we consider the problem of communicating reliably over a MIMO block Rayleigh fading channel in the low SNR regime.   We assume the noncoherent model, wherein neither the transmitter nor the receiver are assumed to have instantaneous channel state information (CSI), while both have knowledge of the channel distribution. In scenarios where the mobile receivers are moving at a high speed or when the number of transmit antennas is large, channel estimation at the receiver might be insufficient due to the small coherence times involved. The problem of the receiver acquiring CSI is further exacerbated in the low SNR regime, where the channel estimates can be unreliable. As a result, the more common assumption of perfect CSI at the receiver, namely that of coherent communications, may not hold true in such cases. 

A more fundamental rationale for studying the noncoherent model is as follows. Since in practice the channel is not known to the receiver at the start of communication, an information theoretic formulation of the noncoherent problem---which implicitly accounts for the resources needed for implicit channel estimation without constraining the transmission scheme in any way---is more fundamental than the coherent formulation. Systems that assume coherent transmission by arguing that the channel can be acquired at the receiver by the use of pilot-symbol assisted transmission to perform explicit channel estimation are inherently suboptimal in general while not taking into account the resources, namely energy and degrees of freedom, needed for pilot transmissions, as they should.

The study of noncoherent fading channels at low SNR is motivated by their application in wideband (WB) and ultra-wideband (UWB) channels. In such scenarios,   the signal power is spread over a large bandwidth, rendering the SNR per degree of freedom low. Transmissions over wideband fading channels experience both time and frequency selectivity. However, within a short window of time or frequency, the channel fading coefficients are known to be highly correlated. One widespread approach therefore to deal with frequency-selectivity is to divide the original wideband channel into several parallel narrowband channels such that each narrowband channel experiences flat fading or a single tap coefficient. 
To deal with time-selectivity, a common approach is to model each narrowband channel through \emph{block fading}. In the block fading model, the channel coefficients are assumed fixed for a duration in time following which they assume independent and identically distributed realizations (here adequate interleaving across time and frequency windows is implicitly assumed). In this work, we model the wideband channel as a block faded narrowband channel in the low SNR regime. This simplifying channel modeling assumption helps captures the essence of the orignal wideband channel, and is widely adopted in the analysis of MIMO fading channels. 

The study of noncoherent SISO fading channels at low SNR dates back to the 1960's. Two equivalent notions of optimality in the literature that are indicators of energy efficiency in the low SNR regime are (1) the input being first order optimal with respect to Shannon capacity or (2) the input achieving the minimum energy per bit or $\frac{E_b}{N_0}_{min}$ required for reliable communication. A classical result by Shannon \cite{ShannonCE:comm} is that in the limit of infinite bandwidth or vanishing SNR, the minimum energy/bit required for reliable communications over an AWGN channel is $-1.59 \mbox{dB}$.  Early work by Kennedy \cite{KennedyRS:Fading}, Jacobs  \cite{JacobsI:incoherent} (also see Gallager \cite{GallagerRG:Info} and the references therein) studied wideband SISO Rayleigh fading channels with an average power constrained input and showed that in the limit of infinite bandwidth or vanishing SNR, the required minimum energy/bit is again $-1.59 \mbox{dB}$, the same as that of an AWGN channel.  A remarkable observation then was that the minimum energy/bit required is the same whether or not the receiver has knowledge of the channel fading coeffecients. Telatar and Tse \cite{TelatarIE:wideband}, and Verdu \cite{VerduS:efficiency} show that the minimum energy/bit is $-1.59$ dB even for fairly general multipath SISO fading channels and general MIMO fading channels, respectively. A common approach  adopted to obtain  $\frac{E_b}{N_0}_{min}$ for fading channels  is to consider the achievable rate of a certain scheme (often M-ary Frequency Shift Keying or MFSK), which is transmitted at arbitrarily low duty cycles (cf. \cite{KennedyRS:Fading,GallagerRG:Info,TelatarIE:wideband}). The required result is then obtained by either showing that the energy/bit of the scheme at vanishing SNR matches that of the AWGN channel, or by deriving an upper bound on capacity that is tight with respect to the achievable lower bound. However, this approach fixes the input {\em a priori}, and therefore  no determination can be made as to the {\em necessary} conditions for a constellation to achieve the minimum energy/bit. The characterization of the class of signals (more generally, input distributions) that are both necessary and sufficient to achieve the minimum energy/bit had been an important and long standing open problem. 

Signals such as arbitrarily low duty-cycled FSK tend to have prohibitively large  peak-to-average-power ratios (PAPR) and are consequently difficult to implement in practice. Such signals are therefore referred to as ``peaky" signals in the literature. Using certain types of fourth moments of the input as measures of peakiness, Medard and Gallager \cite{MedardM:bandwidth}, and Subramanian and Hajek \cite{SubraVG:broadband} showed that signaling that is not peaky in either time or frequency dimensions cannot achieve the minimum energy/bit as ${\rm SNR} \to 0$. Verdu \cite{VerduS:efficiency} formalized this notion further for fairly general noncoherent MIMO fading channels and established that \emph{flash signaling}, where the input distribution converges to a zero mass and a non-zero mass that is transmitted with vanishing probability as ${\rm SNR} \to 0$, is both necessary and sufficient to achieve the minimum energy/bit. While noncoherent communications is sufficient to transmit at the AWGN minimum energy/bit of $-1.59 \mbox{dB}$, the work in \cite{VerduS:efficiency} resolves another major difficulty. It introduces and explains the crucial role of \emph{wideband slope} (${\mathcal{S}}_0$) at large but finite bandwidths. The wideband slope is a measure of how fast the energy/bit of the optimal scheme approaches the minimum energy/bit, and is synonymous with the notion of second order optimality with respect to Shannon capacity. One main result of \cite{VerduS:efficiency} is that for noncoherent MIMO channels with an average power constraint, the wideband slope is zero. This result implies that to approach the minimum energy/bit, the bandwidth for reliable noncoherent communications becomes prohibitively large and the associated signaling scheme prohibitively peaky, and therefore no realistic (i.e., bandwidth limited and peak-limited) scheme can achieve the minimum energy/bit. 

Hence it was important to pose problems that provide meaningful second-order performance when considering noncoherent fading channels at low SNR. One way was to impose suitable peak-constraints on the input. It is shown in Rao and Hassibi \cite{RaoC:lowSNR} that under certain regularity conditions on the signal, which include making the fourth and sixth moments finite, the noncoherent MIMO capacity grows as $O({\rm SNR^2})$.  Similar expressions for the mutual information up to the second order are obtained in closed form in \cite{PrelovV:secondorder , HajekB:smallpeak} with different assumptions on the fading matrices and peak-power constraints. Even though such problems have capacity behaving as $O({\rm SNR^2})$, and hence the minimum energy/bit not occurring at a vanishing SNR, they are important since they involve practical modulation schemes with reasonable PAPR. Schemes designed to satisfy such regularity conditions must be deployed in the vicinity of the SNR where the minimum energy/bit is achieved. Also relevant is the interesting case of the peak-constraint imposed being independent of the average power constraint, resulting in $O({\rm SNR})$ growth of capacity. In this case, it will be shown here that the wideband slope is not zero anymore (unlike the average power constraint only problem). Therefore, the energy/bit approaches the minimum energy/bit at a non-zero rate as ${\rm SNR} \to 0$.
Gursoy and Verdu \cite{GursoyMC:nonc} consider SISO Rician fast fading channels and impose different peak power constraints in addition to the average power constraint on the input.  For certain combinations of peak and average-power constraints, they characterize the $\frac{E_b}{N_0}_{min}$ and ${\mathcal{S}}_0$ for  SISO Rician fast fading channels. 
 For a combination of peak and average power constraints, they show that On-Off Quadrature Phase Shift Keying (OOQPSK) achieves the minimum energy per bit as well as the optimal wideband slope for the noncoherent SISO Rician fast fading channel. This result is obtained in \cite{GursoyMC:nonc} by directly evaluating a second order expansion of mutual information for OOQPSK, and this approach cannot be extended to more general MIMO block fading models. To the best of the authors' knowledge, this is the only input distribution reported in the literature that is both first and second order optimal, in the context of peak-constrained noncoherent communications over fading channels. 

Abou-Faycal et. al. \cite{Abou-FaycalIC:capacity} consider a noncoherent SISO Rayleigh fast fading channel and prove that the capacity achieving distribution is discrete with a finite number of points, one of them being at the origin. In \cite{GursoyMC:capacity}, the authors consider a SISO Rician fast fading channel and show that the capacity achieving distribution is discrete even when certain types of peak-constraints are imposed. While there is no formal proof of the discreteness of the optimal input for MIMO Rayleigh fading channels, it is expected to be the case. Despite these results, discrete input optimization of information theoretic measures is rarely considered since the optimizations encountered are often seen as being analytically intractable. Another compelling reason for considering the problem of maximizing  mutual information as a finite dimensional optimization, over a discrete and finite cardinality input is that, the solution, if obtainable, would offer insights simultaneously into information theoretic as well as coding-modulation aspects. For, consider that even when capacity achieving probability distribution functions are found, the problem of practical transmission would be still unresolved as it would not be clear how the choice of a quantization of the optimum input would affect performance. Some recent works that deal with discrete signal constellation design using information theoretic criteria but only under an average power constraint and via {\em numerical} optimization techniques are \cite{HuangJ:optdist, SrinivSG:const, BorranMJ:NoncSTC}. While the results in \cite{SrinivSG:const} provided numerically computable tight lower bounds on capacity of the noncoherent MIMO channel, the associated constellations may be hard to implement in practice due to their limited analytical structure and lack of strict peak or peak-to-average power ratio constraints.

In this paper, we pose and solve two key problems of obtaining the optimum discrete input of finite cardinality for peak-constrained MIMO noncoherent block Rayleigh fading channels in closed form. Given the results of \cite{Abou-FaycalIC:capacity,GursoyMC:capacity}, it is expected that there will be no loss in optimizing over discrete inputs as opposed to input distribution functions. In both problems, we assume average power constraints on the input. In addition, we also assume natural peak constraints per antenna and per time slot, which closely emulate constraints on power amplifiers, instead of fourth and higher order moment constraints on the input used in \cite{MedardM:bandwidth,SubraVG:broadband,RaoC:lowSNR}.  In the first problem, the peak power to average power ratio (PPAPR) of the input constellation is held fixed, while in the second, the peak power is fixed independently of the average power. We refer to these two problems as the \emph{PPAPR constrained} and \emph{peak-constrained} cases, respectively.  We show that interestingly, in the case of the noncoherent MIMO Rayleigh fading channel at low SNR, such joint optimizations of information theoretic metrics over complex signal matrices and their respective probabilities are indeed analytically tractable and result in elegant closed form solutions. 

In the PPAPR constrained case, it can be shown that the input satisfies certain regularity conditions specified in \cite{RaoC:lowSNR}. For such inputs, the mutual information is obtained up to second order in \cite{RaoC:lowSNR} and shown to grow as $O({\rm SNR}^2)$.  In one of the key contributions here, we maximize this second order mutual information jointly over the matrix-valued elements of a finite input constellation and their probabilities, when the cardinality of the constellations is no greater than $T+1$, where $T$ is the channel coherence blocklength. 

In the peak constrained case, the mutual information behaves as $O({\rm SNR})$. Here, we explicitly characterize the structure of  constellations of any finite cardinality that are optimal up to first order, or equivalently, that minimize energy/bit or maximize capacity per unit energy.  More importantly, among constellations of cardinality no greater than $T+1$ that are first-order optimal, those that maximize the mutual information up to second order, or equivalently, the wideband slope, are characterized. 

In both PPAPR and peak constrained problems, the optimal solutions are obtained in closed-form to finite dimensional non-convex optimizations. Moreover, the solutions are established to be both necessary and sufficient to optimize their respective information theoretic metrics. Interestingly, the solutions to both the PPAPR-constrained and peak-constrained problems are found to be identical. Due to its special structure, we refer to the common solution as Space Time Orthogonal Rank one Modulation, or STORM.  

Moreover, in the PPAPR constrained case, STORM (with cardinality $T+1$) is shown to be near-optimal even among constellations of unconstrained cardinality, even for modest values of $T$ and PAPR. Hence, there is not much to be gained by using more than $T+1$ points in this case. In the peak-constrained case, we first obtain  necessary and sufficient conditions for a constellation of \emph{any} finite cardinality to achieve the minimum energy/bit. Among all such constellations, when the cardinality is no greater than $T+1$, STORM is established as being both first and second order optimal. Our approach provides a far more detailed characterization of the first and second order behavior of noncoherent MIMO capacity than in existing literature. Specifically, we show that when the peak power is less than a certain threshold, it is possible to have a wideband slope that is non-zero, and obtain the maximum wideband slope achievable by a $T+1$ point p.m.f. Moreover, the energy/bit and the wideband slope achieved by STORM reveal a fundamental energy-vs-bandwidth efficiency tradeoff that enable the determination of the operating (low) SNR and peak power most suitable for a given application.
 
It also follows from our analysis and optimization that while the conventional MIMO On-Off Keying (OOK) also achieves the minimum energy per bit, STORM has a wideband slope that is $T$ times greater which translates into an increase in bandwidth efficiency (or a decrease in the PAPR) by a factor of $T$ in the wideband regime. Given typical values of the coherence blocklength $T$, these gains are potentially huge.  Our results and conclusions also temper the conclusions of \cite{VerduS:efficiency} obtained under only the average power constraint regarding noncoherent communications over fading channels.

Among the several new insights that STORM provides on communications in the low SNR regime one that runs contrary to conventional wisdom is that, under the practical constraints considered in this work, it helps to use \emph{all} available transmit antennas, not just one, to transmit linearly dependent signals across them in the low SNR regime. 

Note that in this work, the input distribution is not {\em a priori} assumed or restricted as it is in most prior work. STORM is obtained through novel techniques involving non-convex optimization of information theoretic measures. Consequently, our approach provides necessary and sufficient conditions for a constellation to be optimal for the noncoherent MIMO Rayleigh fading channel, resolving a long-standing open problem. Low duty cycled M-ary FSK (MFSK) \cite{KennedyRS:Fading,GallagerRG:Info,TelatarIE:wideband} which is often proposed to achieve first order optimality in a SISO channel, is seen to be closely related to a special case of STORM. However, the zero symbol in STORM is information bearing which is not the case in low duty cycled MFSK. This can make STORM have higher achievable rates especially in the PPAPR-constrained case. Moreover, in this work, we specify a class of STORM constellations. One subtle insight afforded through different STORM constellations is that optimal signal constellations need not be peaky in frequency dimension (as in low duty-cycled MFSK), in addition to be being peaky in time dimension. In the process, we discover a new optimal SISO constellation which may be called ``Permuted MFSK" due to its relation to MFSK but would have better spectral properties in general.

To close this section, some notational conventions used throughout the paper are described. Matrices are denoted by the boldfaced capital letters, and vectors by bold faced small letters. The symbol $\otimes$ denotes the Kronecker product. The matrices $\Xb^{T}$, $\bar{\Xb}$ and $\Xb^{*}$ denote the transpose, complex-conjugate, and conjugate transpose of $\Xb$, respectively. Moreover, $\tr (\Xb)$ and $\left| \Xb \right| $ denote the trace and determinant of the matrix $\Xb $. The notation $\left[\Xb\right]_{ij}$ refers to the $(i,j)^{th}$ element of the matrix $\Xb$. The notation $\Xb^{(m)}$ refers to the $m^{th}$ row of the matrix $\Xb$. For an integer $N$, $\Ib_N$ is an $N \times N$ identity matrix and ${\bf{1}}_{N}$ is the $N$ length column vector of ones. The block diagonal matrix with matrices $\Ab_1, \dots, \Ab_N$ along the block diagonal and zeros elsewhere is denoted as $\mbox{blockdiag}(\Ab_1, \Ab_2, \dots , \Ab_N)$. $\E[.]$ denotes the expectation operator. A function $f(\rho)$ is said to behave as $o(\rho)$ when $\lim_{\rho \to 0} \ \frac{f(\rho)}{\rho}  = 0$. The symbol $X^{C}$ is used to denote the complement of the set $X$. The symbol $\preceq$ is used to denote generalized inequality ,i.e., if $\Ab \preceq \Bb$ then $\Bb - \Ab$ is positive semidefinite (psd). The first and second derivatives of a function $f(x)$ at $x=c$ are denoted by $\dot{f}(c)$ and $\ddot{f}(c)$, respectively. The function $\log(.)$ always refers to natural logarithm, unless otherwise specified. Complex, circularly symmetric, Gaussian random vectors with mean $\mb$ and covariance matrix $\Qb$ are said to be $\NCplx(\mb,\Qb)$ distributed.

 \section{System Model} \label{STORMsystemmodel}

Consider a MIMO channel with $N_t$ transmit and $N_r$ receive antennas. The random channel matrix $\Hb \in \Cplx^{N_t \times N_r}$ is assumed to be constant for a duration of $T$ symbols after which it changes to an independent value. It has independent, identically distributed (i.i.d.) $\NCplx(0,1)$ entries. The knowledge of the distribution of $\Hb$ is known to the transmitter and receiver. The realizations of $\Hb$ however, are unknown at both ends. With the transmitted symbol denoted as $\Xb \in \Cplx^{T \times N_t}$, the output of the channel can be written as 
 \begin{eqnarray}
 \Yb = \Xb\Hb + \Nb \ .\label{STORMsyseqn1} 
 \end{eqnarray}
 The entries of the additive noise matrix $\Nb$ are assumed to be i.i.d. $\NCplx(0,1)$ distributed random variables. The symbol $\Xb$ is drawn from a finite constellation or alphabet $\Ck$ with matrix-valued elements.  



Two key cases based on the types of constraints imposed are considered in this work.

(i) \emph{PPAPR-constrained case :}
It is assumed that the average SNR at each receive antenna is constrained to be $P$ so that
 \begin{eqnarray}
 \frac{1}{T}\mbox{E}[\tr(\Xb\Xb^{*})]\leq P \, . 
 \label{STORMpowerconstraint}
 \end{eqnarray}
Moreover, a peak-power constraint is imposed per space-time slot, namely,
\begin{eqnarray}
\|\Xb\|_{\infty} \define \max_{i,j} \ |\left[\Xb\right]_{ij}| \leq \sqrt{K} \ \ , \forall \Xb \in \Ck \ .\label{STORMpeak-power}
\end{eqnarray}
This is most natural and  practically meaningful peak-power constraint as it restricts the peak-power per antenna and per time slot (to be at most $K$).  
It accurately models constraints on individual transmit RF power amplifiers in practice. The PPAPR constraint is that the ratio $\frac{K}{P}$ is taken to be a fixed constant. This condition ensures that as the average SNR $P \to 0$, the maximum peak-power also goes to zero.  


(ii) \emph{Peak-constrained case :}
Here, the average power constraint (\ref{STORMpowerconstraint}) and the peak-power constraint (\ref{STORMpeak-power}) are assumed to hold. In this case however, $K$ is  assumed to be a fixed constant {\em independent} of $P$. In other words, in contrast to the PPAPR-constrained case, the peak power remains constrained by $K$ (and does not change) as the average SNR $P \to 0$.

For convenience, we will denote the average energy per block of $T$ symbols as $E = PT$.  

The noncoherent MIMO Rayleigh fading channel thus described is completely specified by the input constraints and the transition probability density function (p.d.f.) of $\Yb$ conditioned on $\Xb$ being transmitted and is easily seen to be
 \begin{eqnarray*}
 p(\Yb | \Xb)= \frac{\exp\left\{-\tr \left(\Yb^{*}\left(\Ib_{T} +  \Xb\Xb^{*}\right)^{-1}\Yb\right)\right\}}{\pi^{TN_r}\left|\Ib_{T} +  \  \Xb\Xb^{*}\right|^{N_r}} \, .
 \end{eqnarray*}

Finally, there will also be occasion to use the notion of the peak-to-average power ratio (PAPR) of a constellation $\Ck$ which is defined as
 \begin{eqnarray} \label{STORMpapr}
 \displaystyle \max_{m, n} \ \max_{\Xb \in \Ck} \ \frac{ \left|\left[\Xb\right]_{mn}\right|^2}{\E\left\{\left|\left[\Xb\right]_{mn}\right|^2\right\}} \ .
 \end{eqnarray}
 
 \section{Maximizing the mutual information at low SNR under the PPAPR constraint}\label{STORMseclowSNRMI}
Consider the above-defined finite input and continuous output noncoherent MIMO Rayleigh fading channel over which the input constellation $\{\Xb_i\}_{i=1}^{L}$ is used with corresponding transmission probabilities $\{P_i\}_{i=1}^L$. The mutual information between the transmitted and received signals, normalized by the block length $T$ (in units of nats/dimension), is thus given as
  \begin{eqnarray} \label{eq:ixy}
  I(\Xb ; \Yb) = \frac{1}{T}\sum_i P_i \int_{\Yb} p\left(\Yb | \Xb_i\right) \log\left(\frac{p\left(\Yb | \Xb_i\right)}{p\left(\Yb\right)}\right) \ d \Yb \, .
  \end{eqnarray}
A closed form expression for $  I(\Xb ; \Yb) $ is unfortunately not known for general SNR. At asymptotically low SNR however, and when the input signal satisfies certain regularity conditions to avoid inputs being prohibitively peaky, the authors in \cite{RaoC:lowSNR} show that the mutual information is zero up to first order for the continuous input and continuous output counterpart of the above channel. Moreover, the mutual information up to the second order in $P$ is also obtained in closed form through a Taylor series expansion and without any assumption on the signal structure beyond the regularity conditions. Note that the expression for mutual information up to second order was also derived  earlier in \cite{PrelovV:secondorder} and \cite{HajekB:smallpeak}, but with more stringent conditions on the input distribution.



For the sake of completeness, the key theorem in \cite{RaoC:lowSNR} for the continuous input and continuous output channel, slightly modified to account for the different power normalizations in this paper, is stated next.
  \begin{theorem} \cite[Theorem 1]{RaoC:lowSNR}\label{STORMRaotheorem} 
  Let $p(\Yb)$ denote the p.d.f. of $\Yb$.

  1. {\em First order result :} If (i) $\frac{\partial p(\Yb)}{\partial P}$ exists at $P=0$, and  \ (ii) $\lim_{P \to 0}  \frac{\E\left[\tr\left\{\left(\Xb\Xb^{*}\right)^2\right\}\right]}{P} = 0$, the mutual information between the transmitted and received signals $\Xb$ and $\Yb$ is zero to first order in $P$, i.e. , $I(\Xb;\Yb)=o(P)$.

  2. {\em Second order result :} If, in addition, (i) $\frac{\partial^2 p(\Yb)}{\partial P^2}$  exists at $P=0$, \ 
(ii) $\E \left[ \tr\left\{\left(\Xb\Xb^{*}\right)^2\right\}\right] < \infty$ \ and \ (iii)  $\lim_{P \to 0}  \ \frac{\E\left[\tr\left\{\left(\Xb\Xb^{*}\right)^3\right\}\right]}{P^2} = 0$, then the mutual information between $\Xb$ and $\Yb$ up to second order in $P$ is given by
  \begin{eqnarray}
  I(\Xb ; \Yb) &=&\frac{N_r}{2T} \tr \left\{ \E[\left(\Xb\Xb^*\right)^2] -  \left(\E[\Xb\Xb^{*}]\right)^2\right\}  + o(P^2)  \ . \label{STORMlowSNRMutInf}
  \end{eqnarray} 
  \end{theorem}
\vspace{3mm}

The applicability of the above result to the discrete input channel with the PPAPR constraint is next discussed. Firstly, following the proof of the above theorem in \cite{RaoC:lowSNR}, it can be seen to hold for the discrete input (and continuous output) case and yield the same expression as in (\ref{STORMlowSNRMutInf}) for mutual information with the expectations in (\ref{STORMlowSNRMutInf}) now over the discrete instead of continuous input as in \cite{RaoC:lowSNR}. The existence of the first and second derivatives of $p(\Yb)$ at $P=0$ are easily verified for the problem at hand. With the PPAPR constraint in effect, the peakiness conditions, namely conditions 1.ii and 2.ii and 2.iii of Theorem \ref{STORMRaotheorem}, are also easily verified to hold as well. Hence, it can be concluded that for a discrete input satisfying the PPAPR constraint (i) the mutual information is zero up to first order in $P$ and (ii) denoting the coefficient of $P^2$ in the mutual information $I(\Xb ; \Yb)$ of (\ref{eq:ixy}) as $I_{\rm low}$, 
\begin{eqnarray}
I_{\rm low}& \define &\lim_{P\to 0} \frac{I(\Xb ; \Yb)}{P^2} = \lim_{P\to 0} \frac{N_r}{2P^2T} \tr \left\{ \E[\left(\Xb\Xb^*\right)^2] -  \left(\E[\Xb\Xb^{*}]\right)^2\right\}\, . \label{STORMlowSNRMutInfsecond}
\end{eqnarray}
Evidently, the dominant second order term in the mutual information at low SNR is $I_{\rm low} P^2$. The problem of interest is hence to maximize $I_{\rm low}$ over $\{\Xb_i\}_{i=1}^{L}$ and $\{P_i\}_{i=1}^L$ under an average power constraint $\sum_i P_i \tr (\Xb_i\Xb_i^{*}) \leq E$ and a peak power constraint  $\|\Xb_i\|_{\infty} = \max_{m,n} \left|\left[\Xb_i\right]_{mn}\right| \leq \sqrt{K} \ \ \forall i$. 

Before unveiling the solution to the above problem, we note that in \cite{RaoC:lowSNR} the mutual information up to second order is maximized over continuous input distributions under two different peak power constraints. The solutions however rely on the assumption that the input signal has the form 
 \begin{eqnarray}
 \Sb&=&\Phib\bar{\Vb} \ , \label{STORMmarzform}
 \end{eqnarray} 
where $\Phib$ is an isotropically distributed unitary random matrix and $\bar{\Vb}$ is a diagonal (random) matrix with non-negative entries.  While this imposition entails no loss of optimality for the case when only the average power is constrained (which is a seminal result of \cite{MarzettaTL:cap}), it does result in a loss of optimality, and a significant one at that, when the peak-power constraint of \cite{RaoC:lowSNR} is enforced which is that the diagonal entries $|\bar{V}_i|^{2} \leq K$. Due to the suboptimal restriction in (\ref{STORMmarzform}), the maximizations in \cite{RaoC:lowSNR} lead to the misleading conclusion that it is optimal to use a single transmit antenna in the low SNR regime. In \cite{HajekB:smallpeak} also, the authors perform the same maximization over continuous input distributions but under a more relaxed peak-constraint $\tr(\Xb\Xb^{*}) \leq \epsilon$ and conclude that a single antenna should be used. Different from \cite{RaoC:lowSNR} and \cite{HajekB:smallpeak}, the optimization problem considered here does not sub-optimally restrict the signals to be as in (\ref{STORMmarzform}) while considering averaged power constrained discrete inputs and the practically relevant peak-power constraint per space-time slot. These assumptions result in a significantly different and more challenging problem than those considered in \cite{RaoC:lowSNR} or \cite{HajekB:smallpeak}.
Indeed, in contrast to \cite{RaoC:lowSNR} or \cite{HajekB:smallpeak}, our results indicate that in the PPAPR-constrained problem, at sufficiently low SNR, it actually helps to use \emph{all} transmit antennas.

For the PPAPR-constrained problem, the set of all feasible constellations with cardinality $L$  is denoted as $\Sk_L$ and can be described as
 \begin{eqnarray*}
\Sk_L = \left\{\left(\Xb_i \ , P_i\right)_{i=1}^{L} \ : \  P_i \geq 0  \ , \ \Xb_i \in \Cplx^{T \times N_t} \ , \ \sum_{i=1}^{L} P_i = 1, 
  \sum_{i=1}^{L} P_i \ \tr(\Xb_i \Xb_i^*) \leq E  , \|\Xb_i\|_{\infty}  \leq \sqrt{K}, \ \forall i \right\} \, .
 \end{eqnarray*}
It is assumed, without loss of generality, that $K N_t T \geq E$, because otherwise, the average power constraint cannot be active and one can therefore solve the problem by changing the average power constraint to $E' = K N_t T$. Let the PPAPR be denoted as $\zeta = \frac{K}{P}$, a constant in the PPAPR-constrained case as $P$ varies.

Let $I_{{\rm low} , L}^{*}$ be the maximum  mutual information up to second order achievable by any constellation in the set $\Sk_L$, so that  \begin{eqnarray}
 I_{{\rm low}, L}^{*} &=&\displaystyle \max_{\left(\Xb_i \ , P_i\right)_{i=1}^{L} \in \Sk_L} \ \ I_{\rm low}\ . 
  \end{eqnarray}
Note that when $P_i=0$, the symbol $\Xb_i$ is not used and therefore the set of feasible constellations in $\Sk_{L'}$ is included in the set $\Sk_{L}$ for any $L' < L$. Hence, $I_{{\rm low}, L}^{*} $ is the maximum mutual information up to second order achievable by any constellation of cardinality no greater than $L$.
The maximum mutual information up to second order when there is no upper limit on the cardinality of the discrete input constellation is defined as $I_{\rm low}^{*} = \lim_{L \to \infty} I_{{\rm low}, L}^{*}$. It will be shown in what is to follow that $\left.I_{{\rm low}, L}^{*}\right|_{L=T+1}$ (and its associated constellation of size $T+1$) is near-optimal in that it can be very close to $I_{\rm low}^{*} $ (and the as yet unknown constellation which achieves the latter).

 The following theorem is one of the main results in this paper.



 \begin{theorem} \emph{(PPAPR-constrained case)} 
 \label{STORMmaxMIlowlarge} Let the coherence time $T \geq 2$. When $L \leq T+1$, the maximum second order mutual information with an $L$-point input constellation is given as
\begin{eqnarray}
I_{{\rm low}, L}^{*}&=&\frac{N_tN_rT}{2}\left(\zeta - \frac{1}{(L-1)N_t}\right) \ .
\end{eqnarray}
An  $L$ point constellation (or p.m.f.) achieves $I_{{\rm low}, L}^{*}$ with $L \leq T+1$ 
if and only if (iff) it is of the following form
 \begin{eqnarray}
 (\Xb_i \ ,  P_i)&=& \left(\sqrt{K} \ \vb_i \wb_i^{*} \ , \ \frac{E}{(L-1)K N_t T}\right)   \ \ , \ \ 1 \leq i \leq L-1 \label{STORMnonzeropoints} \\
 (\Xb_L \ , P_L)&=& \left(\Zrb_{T \times N_t} \ , \ 1 - \frac{E}{K N_t T}\right) \ , \label{STORMlargepeakopt}
 \end{eqnarray}
where for each $i$, $\vb_i \in \Cplx^{T \times 1}$ is the $i^{th}$ column of a unitary matrix $\Vb$, $\wb_i \in \Cplx^{N_t \times 1}$ and 
\begin{eqnarray} \label{eq-vwcons}
|\left[\vb_i \wb_i^{*}\right]_{mn}| = 1, \ \forall \ i, m, n \ .
\end{eqnarray}
Furthermore, $I_{\rm low}^{*}$, the maximum second order mutual information with an unconstrained cardinality, is bounded above and below as 
 \begin{eqnarray}
   \left.I_{low , L}^{*}\right|_{L=T+1} = \frac{N_r}{2}\left( \zeta N_t T - 1\right)   \ \leq \  I_{low}^{*} \ < \frac{N_r}{2} \zeta N_t T \ . \label{STORMMIlargepeak} 
 \end{eqnarray}
 \end{theorem}

\proof The proof is given in Section \ref{STORMTheoremproofs}.
\endproof

The optimal signal constellation for $L=T+1$ given in Theorem \ref{STORMmaxMIlowlarge} can be viewed as a space-time code (employing unequal transmission probabilities) that achieves the maximum mutual information up to second order at low SNR. Based on its structure, it is referred to as  {\bf S}pace {\bf T}ime {\bf O}rthogonal {\bf R}ank one {\bf M}odulation (STORM)  because each non-zero matrix is of  unit rank and is orthogonal to the other constellation matrices by construction.
 Two examples of matrices that can be used for the unitary matrix $\Vb$ are the Discrete Fourier Transform (DFT) matrix and the Hadamard matrix (when it exists). In one embodiment of STORM, $\wb_i$ can be chosen to be ${\bf 1}_{N_t} \, \; \forall i$. In this case, each of the $L-1=T$ non-zero constellation points is formed from a column of $\Vb$ and this column is repeated over the $N_t$ antennas. The $L^{th}$ point is of course the all-zero matrix.

 It can be seen that for STORM, the PAPR as defined in  (\ref{STORMpapr}), is $\frac{KN_t}{P} =\zeta N_t \geq 1$. 
Clearly, the ratio between the upper and lower bounds on $I_{\rm low}^*$ in (\ref{STORMMIlargepeak}) is nearly equal to unity when $\zeta N_t T >> 1$. This is evidently true even for moderate and practical values of PAPR and $T$.  As an example, for $\zeta = 2$, $N_t=2$ and $T=4$, the ratio is $0.94$. Hence, even for moderate values of PAPR and $T$,  the $T+1$ point STORM almost achieves  $I_{\rm low}^*$ (the limit with unconstrained cardinality) and there is not much to be gained by using more than $T+1$ points. 


\subsection{Remarks} \label{Remarks1}

Since STORM achieves a significant fraction of $I_{\rm low}^{*}$  even for moderate values of $\eta$ and $T$,  the following insights from its structure and mutual information up to second order it achieves at low SNR are of interest. For brevity, the mutual information up to second order at low SNR is simply referred to as mutual information in the rest of this section.


\begin{enumerate}
\item{}  It can be seen that the mutual information of STORM increases linearly with the maximum peak power $K$. That it is an increasing function is to be expected since peaky signaling is known to achieve  the noncoherent capacity in the low SNR regime when there is only an average power constraint. Moreover, the mutual information also increases linearly as a {\em product} $N_t . N_r $ of the numbers of transmit and receive antennas. The use of a single antenna is evidently suboptimal by a factor of $N_t$.

\item{} A reason that is often cited in the literature for explaining the efficacy of using a single antenna at low SNR is that the number of channel parameters that are to be implicitly estimated is the least in this case. The use of a single antenna however is not necessary to ensure this and can even be detrimental to performance as explained above. Consider STORM, where the received signal when the $i^{th}$ non-zero signal is transmitted is  \begin{eqnarray}
 \Yb = \sqrt{K} \ \vb_i \wb_i^{*} \Hb + \Nb  
  = \sqrt{K} \ \vb_i \hb^{T} + \Nb \ , \label{STORMimplicitest}
 \end{eqnarray}
where $\hb^{T} =  \wb_i^{*} \Hb$ and so $\hb$ is $\NCplx(\Zrb , N_t  \Ib_{N_r})$ distributed. Therefore, the {\em effective} channel (\ref{STORMimplicitest}) does in fact involve only $N_r$ (and not $N_tN_r$) unknown channel coefficients even though all transmit antennas are used. The optimality of the unit rank structure of STORM could thus be indeed attributed to the difficulty of (implicit) estimation of $N_tN_r$ coefficients  at low SNR because it avoids this task by focusing the power on just $N_r$ effective unknown path gains, while at the same time making use of all the transmit antennas. 

\item{}  Consider the case when $\wb_i = {\bf 1}_{N_t} \ \forall i$ in  (\ref{STORMnonzeropoints}), which is sufficient for $T+1$ point STORM to be optimal. Then the symbols sent by all transmit antennas at any given time are identical and the fading gains effectively add up at each receive antenna. So, why not just use a single transmit antenna? 
All transmit antennas must be used because otherwise the effective received power is smaller due to the peak-power constraint which limits the symbol power per antenna and per time slot.

\item{}  A canonical embodiment of STORM is one that results from setting $\wb_i = {\bf 1}_{N_t} \, , \, \forall i$ and $\Vb=[\vb_1 \cdots \vb_T]$ to be a $T$-dimensional DFT matrix in  (\ref{STORMnonzeropoints}).
A convenient feature in this DFT version of STORM is that the entries of the signal matrices can be transmitted using PSK symbols with an additional zero point. Alternatively,  a $T$-dimensional Hadamard matrix can be used for $\Vb$ (when it exists).  The advantage of using a Hadamard matrix  is that it is enough to transmit real symbols for {\em each} entry, specifically, BPSK and an additional zero point. Hadamard matrices of dimension $T$ exist when $T = 2^{n}$ for any natural number $n$ and also for many multiples of 4. In Appendix\ref{STORMFFT}, we show how block decoding of STORM may be simplified using either the Fast Fourier Transform (FFT) or the Fast Hadamard Transform (FHT), when $L-1$ is a power of $2$.

\item{} Consider the special case when there is only a peak-constraint on the input (i.e., $KN_tT = E$). Here, it can be seen that STORM has no zero point (so $L = T$) and is given by
 \begin{eqnarray}
 (\Xb_i \ , P_i) = \left(\sqrt{K} \ \vb_i \wb_i^{*}, \frac{1}{L}\right) , \ \ 1 \leq i \leq L \ .     \label{STORMsmallpeakopt} 
 \end{eqnarray}
Hence, all points are equiprobable and the PAPR is unity, thus facilitating practical implementation. Moreover, this constellation is near-optimal when there is only a peak constraint and when $T >> 1$ as seen from the bounds on $I_{\rm low}^* $ in (\ref{STORMMIlargepeak}) of Theorem \ref{STORMmaxMIlowlarge}.

\item{} The canonical version of STORM can be seen as a form of generalized $(T+1)$-ary ON-OFF signaling with repetition coding across the transmit antennas and with unequal probabilities of ON and OFF signaling, with the ON signaling actually being the classical T-ary, equiprobable Frequency Shift Keying (T-FSK). The larger the allowed PPAPR, the higher the probability of the OFF signal. In fact, STORM takes advantage of all the peak power allowed for each space-time slot when transmitting non-zero symbols while meeting the average power constraint by the inclusion of the zero symbol with as high a probability as the PPAPR constraint would allow.

\item{} Consider the special case of a SISO system when there is only a per time-slot peak power constraint $K$. Here, Theorem \ref{STORMmaxMIlowlarge} establishes the second order optimality of equiprobable T-FSK at low SNR among all $T$-ary constellations, and the near-optimality under unconstrained cardinality when $T >>1$. In general for SISO systems however, depending on the peak and average power constraints, an additional zero signal is needed of probability different from that of the $T$ (equi-probable) T-FSK signals.
\item{}  The mutual information of STORM may be expressed as $\frac{EN_r}{2}(KN_t - \frac{E}{T^2})$. For fixed $K$, $T$, and $E$, it increases linearly with $N_t$. This may be attributed to the fact that increasing $N_t$ with fixed $K$, $T$ and $E$ increases the overall peak-power $\tr(\Xb\Xb^{*}) = \zeta N_t E$, while simultaneously decreasing the probability of transmitting a non-zero signal $\frac{1}{\zeta N_t}$, thereby making the signals more peaky in the time domain. On the other hand, when $T$ is increased for a fixed $N_t$ and $E$,  the overall peak-power $\tr(\Xb\Xb^{*}) = \zeta N_t E$ and the probability of transmitting the zero signal $\frac{1}{\zeta N_t}$ remain fixed but the mutual information increases with $T$. To get some insight on why this is so, consider the canonical version of STORM. An increase in $T$ implies that the T-FSK transmissions (repeated over each antenna) become more peaky in the frequency domain \footnote{This was pointed out by a reviewer.}. 

\item{} STORM constellations other than the canonical ones can also be constructed. For example, one can use the inverses of the DFT and Hadamard matrices for a choice of $\Vb$. More generally, if $\widetilde{\Vb}$ is unitary with unit-magnitude elements so is $ \Vb  = \Pb \widetilde{\Vb} \Qb $ where $\Pb$ and $\Qb$ are $T \times T$ permutation matrices. $\Qb$  only permutes the columns of $ \widetilde{\Vb} $ thereby renumbering the signals leaving the STORM constellation unchanged. However, row permutations induced by $\Pb$ would result in constellations that are no longer peaky in the frequency domain as compared to the canonical DFT version of STORM. It is unclear as to how the complete class of STORM constellations can be constructed. In this regard, note that the $\wb_i$ vectors can be arbitrary as long as its elements have unit magnitudes. So ``repetition" across transmit antennas can involve arbitrary phase rotations or multiplication by possibly distinct unit-magnitude complex numbers.

\item{}   The cutoff rate for the discrete input (of cardinality $L$) and continuous output channel is given by 
\begin{eqnarray}
R_0 &=& \max_{\{P_i\}_{i=1}^L , \{\Xb_i\}_{i=1}^{L}} - \log \left\{ \sum_{i}\sum_{j} P_i P_j \int \sqrt{p(\yb / i)p(\yb / j)} d\yb \right\}.  \label{STORMcutoffrate}
\end{eqnarray}
The cutoff rate was initially advocated as a design criterion for modulation schemes in \cite{WozenJM:moddemod} and \cite{MasseyJL}. It is a lower bound on the random coding exponent, and also provides an exponentially accurate description of the attainable error probability when communicating at the critical rate \cite{WozenJM:moddemod}.  Let the argument of $\max(.)$ in (\ref{STORMcutoffrate}) be denoted as the \emph{cutoff rate expression}. For the noncoherent MIMO channel at low SNR, the cutoff rate expression is easily shown to be (c.f. \cite{HeroAO:sig}) 
\begin{eqnarray}
  CR_{\rm low}&=&\frac{ N_r}{8}  \sum_{i,j} P_i P_j \tr \left\{ \left(\Xb_i\Xb_{i}^{*}-\Xb_j\Xb_{j}^{*}\right)^2\right\} + o(P^2)
\end{eqnarray}
An interesting property of $CR_{\rm low}$ \cite{SrinivSG:const} is that when the input constellation satisfies the regularity conditions, $ \lim_{P \to 0} \frac{ CR_{\rm low}}{I_{\rm low}} =  \frac{1}{2} $. In the limit of low SNR therefore, ${\rm CR}$ behaves identically to the mutual information. Therefore, the $T+1$ point STORM also maximizes the cutoff rate expression up to second order at low SNR. 



\item{}  An often used noncoherent constellation design criterion (cf. \cite{AgrawalD:grassman,HochwaldB:sptimesig}) is to maximize the worst-case chordal distance which is given by $\min_{j \neq i} \ \tr \left\{\Ib - \Xb_i^{*}\Xb_j\Xb_j^{*}\Xb_i \right\}$. For STORM, the worst-case chordal distance is the maximum possible as for every $i \neq j$,  $\Xb_i^{*}\Xb_j = \Zrb_{N_t \times N_t}$. Moreover, the  difference between any two distinct matrices in STORM  has unit rank, and hence the scheme would have a diversity order of $N_r$ at high SNR if employed as a coherent space-time code \cite{TarokhV:sptime} whereas constellation design at high SNR for the coherent MIMO channel is typically geared towards achieving maximum diversity ($N_t N_r$). Theorem \ref{STORMmaxMIlowlarge} shows that optimal noncoherent constellations at low SNR have quite the opposite properties from good coherent constellations at high SNR.

\item{} Subsequent to the conference version of this paper \cite{SGSrinivasan:STORMaller} (see also \cite{SGSrinivasan:STORMisit}), Sethuraman et. al. \cite{SethuramanV:MIMO} consider a MIMO  Rayleigh fading channel with the noncoherent assumption and with the fading process modeled as stationary and ergodic, as well as correlated over time. The authors characterize input distributions which are optimal for the stationary and ergodic MIMO channel, under average-power constraints and peak-constraints which are per space-time slot similar to the PPAPR-constrained case here. Interestingly, one  distribution identified in \cite{SethuramanV:MIMO}  which achieves the capacity up to second order can be seen to be closely related to the canonical version of STORM here. While this distribution is  obtained for a different fading process, the channel coherence time $T$ here can be thought of as playing the same role as channel memory in \cite{SethuramanV:MIMO}.

\end{enumerate}

\subsection{Proof of Theorem \ref{STORMmaxMIlowlarge} } \label{STORMTheoremproofs}
In this subsection, the proof of Theorem \ref{STORMmaxMIlowlarge} is given. 
The following definitions and lemmas are needed first from \cite{HorstR:glo}.
  \begin{definition}
  A \emph{convex maximization} problem is an optimization problem in the following form :
  \vspace{-2mm}
  \begin{eqnarray}
  \max_{\xb \ \in \ X} \ \  f(\xb) \ ,
  \end{eqnarray}
  where $f(\xb)$ is a convex function and  $ X \ \subset \ \Re^{n}$ is a convex set.
  \end{definition} 

  \begin{definition}
  A point $\xb$ on the boundary of a convex set $X$ is called an extreme point if there are no distinct points $\xb_1 , \xb_2 \in X$ such that $\xb = \lambda\xb_1 + (1-\lambda)\xb_2 \ , \ 0 < \lambda < 1$.
  \end{definition}
     \begin{lemma}
     A closed, bounded convex set in $\Re^{n}$ is the convex hull of its extreme points. \label{STORMhull}
     \end{lemma}

  \begin{lemma}
  The global maximum of a convex function $f$ over a compact convex set $X$ is attained at an extreme point of $X$. A point in a compact convex set $X$ is a global maximizer of a strictly convex function $f$    iff it is an extreme point of $X$.  \label{STORMextreme}
  \end{lemma}


  \begin{definition} \label{STORMpolytope}
  A \emph{polyhedron} is defined to be the set of points $\Pk = \{\xb \in \Re^{n} : \Ab\xb \leq \bb$, where $\Ab \in \Re^{m \times n}$ and $\bb \in \Re^{m}$. A bounded polyhedron is called a polytope. The extreme points of a polytope are referred to as \emph{vertices}.
  \end{definition}
  
The next lemma gives the necessary and sufficient conditions for a point to be a vertex of a general polytope.
  \begin{lemma} \label{STORMvertex}
With the same notation as in Definition \ref{STORMpolytope}, let $\ab_i^{T} , 1 \leq i \leq m$ denote the rows of the matrix $\Ab$. Further, for $\xb \in \Pk$, let $I = \left\{i \in \{1, \dots, m\} : \ab_i^{T}\xb=b_i\right\}$ describe the inequalities which are binding (active) at $\xb$, and let $\Ab_{I}$ be the matrix with rows $\ab_i^{T}, i \in I$. Then $\xb \in \Pk$ is a vertex of $\Pk$ iff $\mbox{rank}\left(\Ab_{I}\right) = n$.
  \end{lemma}

The following lemma more sharply specifies the vertices of a special polytope which will be useful in the proof of Theorem \ref{STORMmaxMIlowlarge}.
\begin{lemma} \label{STORMvertexchar}
Consider the polytope defined by
\begin{eqnarray}
\Dk&=&\left\{\db :  \sum_i P_i \ d_i \leq E \ , \ 0 \leq d_i \leq Q \ , \  i = 1 , \dots , L \right\} \ ,
\end{eqnarray}
which is the intersection of the half-plane $\sum_i P_i \ d_i \leq E$ and the  hyper-cube $0 \leq d_i \leq Q$. Each vertex of $\Dk$ consists of $L-1$ entries that are either $Q$ or $0$, and exactly one entry $c$ such that $0 \leq c \leq Q$.
\end{lemma}
\proof The polytope $\Dk$ can be expressed in the standard form $\Ab \db \leq \bb$ given in  Definition \ref{STORMpolytope}, by setting
  \begin{eqnarray}
  \Ab = \left[\begin{array}{cccc}
  \qb^T \\
   \Ib_L \\
  -\Ib_L
  \end{array}\right]_{(2L+1) \times L} \qquad {\rm and} \qquad
  \bb = \left[\begin{array}{cccc}
  E \\
  Q  \ {\bf{1}}_L  \\
  \Zrb_L
  \end{array}\right] \ \ \  \ \ \ ,
  \end{eqnarray}
where $\qb = [P_1 \ P_2 \ \dots \ P_L]^T$. Let $\xb$ be a vertex of the polytope described by $\Ab \db \leq \bb$. Then, the rows of $\Ab$ which satisfy $\ab_i^{T}\xb = b_i$ should form a matrix with rank $L$ by Lemma \ref{STORMvertex}.  If $\xb $ is a vertex for which $\qb^T \xb = E$ then there are at least $L-1$ more linearly independent rows of $\Ab$ that correspond to active constraints. Suppose $k$ of them are of the form $x_j = Q$ for $ j \in J \subseteq \{1, 2, \dots, L\}$, then at least $L-1-k$ active constraints (out of the remaining $L-k$ constraints) must be of the form $x_j=0$ for $j \in J^{C}$. Hence, at most one entry of $\xb$ can lie anywhere between $0$ and $Q$ (call it $c$). If $\xb $ is such that $\qb^T \xb < E$, then of course it is a vertex by Lemma \ref{STORMvertex} iff $x_j = Q$ for all $j$ in the subset $J \subseteq \{1, 2, \dots, L\}$ for which $ \sum_{j\in J} P_j < E $ and $x_j=0$ for all $ j \in J^C$ (there are as many such vertices as there are subsets $J$ for which $ \sum_{j\in J} P_j < E $). In this case, all the entires of the vertex are either $Q$ or $0$ (set $c=0$ or $Q$).
\endproof

\proof (of Theorem \ref{STORMmaxMIlowlarge}): The problem that needs to be solved here is essentially
  \begin{eqnarray}
  &\max_{\{P_i\}_{i=1}^{L}, \{\Xb_i\}_{i=1}^{L}}& \ I_{\rm low} \\
   &\mbox{subject to}& \sum_{i} P_i \tr(\Xb_i\Xb_i^{*}) \leq E \nonumber\\ 
   && \|\Xb_i\|_{\infty} \leq \sqrt{K}, \ \forall i \nonumber\\
  && \ \sum_i P_i = 1, P_i \geq 0  \ \forall i \nonumber
  \end{eqnarray} 
  where $I_{\rm low}$ is given in (\ref{STORMlowSNRMutInfsecond}). 
Maximizing $I_{\rm low}$ is equivalent to maximizing
  \begin{eqnarray}
  & &  \sum_i P_i \tr\left(\Xb_i\Xb_i^{*}\Xb_i\Xb_i^{*}\right)  - \tr\left(\sum_i P_i \Xb_i\Xb_i^{*}\sum_j P_j \Xb_j\Xb_j^{*}\right)\\
  &=& \sum_i P_i\left(1-P_i\right)\tr \left(\Xb_i\Xb_i^{*}\Xb_i\Xb_i^{*}\right) - \sum_{i, j\neq i} P_i P_j \tr\left(\Xb_j^{*}\Xb_i\Xb_i^{*}\Xb_j\right)\label{STORMprenegzero}\\
  &\leq& \sum_{i=1}^{L} P_i\left(1-P_i\right)\tr \left(\Xb_i\Xb_i^{*}\Xb_i\Xb_i^{*}\right) \ . \label{STORMnegzero}
  \end{eqnarray}
  Since terms of the form $\tr\left(\Xb_j^{*}\Xb_i\Xb_i^{*}\Xb_j\right)$ are non-negative,  (\ref{STORMnegzero}) follows by replacing all negative terms in (\ref{STORMprenegzero}) by zero. Let $\xb_{ik}$ denote the $k^{th}$ column of the matrix $\Xb_i$. The equality in (\ref{STORMnegzero}) occurs iff $\xb_{jk}^{*}\xb_{il} = 0 \ \forall k, l, j \neq i$. The strategy is to maximize the bound in (\ref{STORMnegzero}) and show later that the signal constellation that maximizes it achieves equality in the inequality in (\ref{STORMnegzero}) when $L \leq T+1$, thereby maximizing $I_{\rm low}$ in these cases. So, let us consider the optimization problem 
  \begin{eqnarray}
  &\displaystyle \max_{\{P_i\}_{i=1}^{L}, \{\Xb_i\}_{i=1}^{L}}& \sum_i P_i(1-P_i)\tr \left(\Xb_i\Xb_i^{*}\Xb_i\Xb_i^{*}\right) \label{STORMoptproblem} \\
  &\mbox{subject to}& \sum_{i} P_i \tr(\Xb_i\Xb_i^{*}) \leq E \nonumber \\
   && \|\Xb_i\|_{\infty} \leq \sqrt{K}, \ \forall i \nonumber\\
  && \ \sum_i P_i = 1, P_i \geq 0  \ \forall i \nonumber
  \end{eqnarray}
 In Appendix\ref{STORMnonconvexity}, a simple argument is given that shows that the maximization of (\ref{STORMoptproblem}) is a non-convex optimization problem. A two-stage approach is thus adopted for solving the optimization in (\ref{STORMoptproblem}). In the first stage, the objective function is maximized over $\{\Xb_i\}_{i=1}^{L}$ while holding $\{P_i\}_{i=1}^{L}$ fixed. In the second stage, the resulting objective function is maximized over $\{P_i\}_{i=1}^{L}$. Furthermore, it is shown that the optimization in the first stage can be split into two successive convex maximization problems and the optimization in the second stage is a convex minimization problem. It is this nice structure that is exploited to obtain the signal matrices $\{\Xb_i\}_{i=1}^{L}$ and the probabilities $\{P_i\}_{i=1}^{L}$ that jointly optimize the upper bound on mutual information (up to second order at low SNR) in (\ref{STORMoptproblem}).

Consider first the optimization in (\ref{STORMoptproblem}) over $\{\Xb_i\}_{i=1}^{L}$ for fixed $\{P_i\}_{i=1}^{L}$. This problem is decomposed into two steps. In the first step, $\tr (\Xb_i\Xb_i^{*}) = d_i$ is fixed for some $ \{ d_i \}_{i = 1}^L$ and the best set of $\{\Xb_i\}_{i=1}^{L}$ is found. Note that $d_i$ is equal to the energy of the $i^{th}$ signal and because of the peakpower constraint, it is sufficient to restrict $d_i \in [0,KN_tT]$. In the second step, the resulting objective function is optimized over $d_i, \, i = 1, \dots, L$.  Geometrically, we first find the matrices $\{\Xb_i\}_{i=1}^{L}$ that maximize the objective function over the contour $\tr (\Xb_i\Xb_i^{*}) = d_i \ \ \forall i$ and then optimize the resulting objective over $\{d_i\}_{i=1}^{L}$, thereby  obtaining the best contour for an arbitrary but fixed $\{P_i\}_{i=1}^{L}$. As it is shown below, both these problems can be solved as convex maximization problems.

With $\tr(\Xb_i\Xb_i^{*}) = d_i \in [0,KN_tT], \ \forall i$, it is clear that the objective function in (\ref{STORMoptproblem}) is maximized when for each $i$, $\Xb_i$ is chosen according to 
  \begin{eqnarray}
  \max_{\tr(\Xb_i\Xb_i^{*}) = d_i \atop{ \|\Xb_i\|_{\infty} \leq \sqrt{K}, \ \forall i}}  \tr \left(\Xb_i\Xb_i^{*}\Xb_i\Xb_i^{*}\right) \ . \label{STORMsubproblem}
  \end{eqnarray}
  Let the eigenvalues of the positive semidefinite matrix $\Xb_i\Xb_i^{*}$ be $\{\lambda_m\}_{m=1}^{T}$ (the dependence on $i$ is implicit). Then, the solution of (\ref{STORMsubproblem}) is upper bounded by the solution of 
  \begin{eqnarray}
  \max_{\sum_m \lambda_m = d_i  \atop {\lambda_m \geq 0, \ \forall m}} \ \sum_m \lambda_m^{2} \ ,  \label{STORMeigsubproblem}
  \end{eqnarray} 
  with equality iff the additional constraints $\|\Xb_i\|_{\infty} \leq \sqrt{K}$ hold for each $i$ for the matrix that achieves the maximum in (\ref{STORMeigsubproblem}). Since the objective function in (\ref{STORMeigsubproblem}) is strictly convex while the constraint set is a polytope, the problem in (\ref{STORMeigsubproblem}) is a strictly convex maximization problem. Hence by Lemma \ref{STORMextreme}, a  solution is globally optimal iff it is a vertex of the constraint set. In this case, the constraint polytope has $T + 1$ vertices which can be found by inspection to be 
  \begin{eqnarray}
  [0 \ \ 0 \ \ \dots \ \ 0 \ \ 0]^{T} \hspace{3mm}, \hspace{3mm}[d_i \ \ 0 \ \  \dots \ \ 0 \ \ 0]^{T} \hspace{3mm}, \hspace{3mm} [0 \ \ d_i \ \ 0 \ \ \dots \ \ 0]^{T} \hspace{3mm}, \hspace{3mm} \dots \hspace{3mm}, \hspace{3mm} [0 \ \ \dots \ \ 0 \ \ 0 \ \ d_i]^{T}  \label{STORMvertices}
  \end{eqnarray}   
since none of them can be expressed as a convex combination of any other points in the set, and any  point in the set can be expressed as a convex combination of the points in (\ref{STORMvertices}). Now, since all the vertices except the all-zero vector give the same value $d_i^2$ for the objective function, $d_i^2$ is the sought maximum. This in turn implies that all the matrices $\{\Xb_i\}_{i=1}^{L}$ have to be of {\em unit rank} for the objective functions to  achieve their maximum value of $d_i^2$ for each $i$ (we adopt the convention that the all zero matrix is of unit rank). Let the number of matrices in $\{\Xb_i\}_{i=1}^{L}$ which are not $\Zrb_{T \times N_t}$ be $L'$. If more than one of the $d_i$'s are zero, they would all correspond to the same zero signal point $\Zrb_{T \times N_t}$ and their respective probabilities would simply add up, resulting in one effective zero symbol matrix.  Therefore,  $L=L'+1$ or $L=L'$ depending on whether or not there is a zero symbol. 

When $L' \leq T$, consider the following constellation
 $\{\Xb_i\}_{i=1}^{L}$,
  \begin{eqnarray}
   \Xb_i&=&\sqrt{\frac{d_i}{TN_t}} \ \vb_i \wb_i^{*} \ \ \ , \ \  d_i > 0 \label{STORMdftconst1} \\
   \Xb_i&=&\Zrb  \hspace{22.5mm} , \ \   d_i=0 \, , \label{STORMdftconst2}
   \end{eqnarray}
   where the vectors $\vb_i$ and $\wb_i$ are constrained as in 
   (\ref{eq-vwcons}).
Note that the set of matrices in (\ref{STORMdftconst1}, \ref{STORMdftconst2}) are of unit rank and satisfy $\tr\left(\Xb_i\Xb_i^{*}\right)=d_i \ , \  \ 1 \leq i \leq T$. Hence they solve the problem in (\ref{STORMeigsubproblem}). Now, since $d_i \leq KN_tT$, using (\ref{eq-vwcons}), it follows that  $\|\Xb_i\|_{\infty} \leq \sqrt{K} \; \forall i$ and hence they also solve the problem in (\ref{STORMsubproblem}).  Moreover, since $L' \leq T$, any pair of different constellation matrices have orthogonal columns (since $\vb_i$'s are orthogonal), which ensures that (\ref{STORMnegzero}) holds with equality. It will eventually be shown that the optimal values of the non-zero $\{d_i\}$ are all equal with $d_i = KN_tT \; \forall i$. This in turn implies that the structure in (\ref{STORMdftconst1}) and (\ref{STORMdftconst2}) is also necessary.

When $L' > T$, the set of $\vb_i$ in  (\ref{STORMdftconst1})  can no longer be selected to be orthogonal to each other. Nevertheless, a set of rank one matrices with the structure given in (\ref{STORMdftconst1}) but with a non-orthogonal set of $\vb_i$ (normalized in the same way),  still solves both (\ref{STORMsubproblem}) and (\ref{STORMeigsubproblem}). Therefore, the expression  $\frac{N_r}{2 T} \sum_{i=1}^{L} P_i(1-P_i) \ d_i^2$ serves as an upper bound on the maximum mutual information up to second order achievable by any constellation of cardinality of $L=L'+1$, which is $I_{low, L}^{*}$. 

In summary the best constellation $\{\Xb_i\}_{i=1}^{L}$ can be specified for any set of non-negative $\{d_i\}_{i=1}^{L}$. It remains to find the best $\{d_i\}_{i=1}^{L}$ according to
  \begin{eqnarray}
  & \displaystyle \max_{\{d_i\}_{i=1}^{L}} & \sum_i P_i(1-P_i) \ d_i^2. \label{STORMsimplifiedmax}\\
  &\mbox{subject to}& \sum_i P_i \ d_i \leq E  \label{STORMhalfplane}\\
  & &0 \leq d_i \leq KN_tT \ \ \forall i \label{STORMhypercube}
  \end{eqnarray}
  For a fixed $\{P_i\}_{i=1}^{L}$, this is also a strictly convex maximization problem over a polytope. Hence, a vertex of the polytope is both necessary and sufficient to achieve the global optimum. The polytope constraint set is exactly of the form considered in Lemma \ref{STORMvertexchar} which states that each vertex would consist of $L-1$ entries that are either $K N_tT$ or $0$, and at most one entry $c$ such that $0 < c < KN_tT$. For vertices for which $ \sum_i P_i \ d_i < E $, it is necessarily the case that all entries are either $0$ or $KN_tT$.



Consider the second stage of the optimization which is over $\{P_i\}_{i=1}^{L}$.
Following the result of the optimization in the first stage, the structure of the optimal $\db$ and the corresponding probabilities are of the form
 \begin{eqnarray}
  \db&=& \left[\underbrace{KN_tT \  \ \dots \  \ KN_tT}_{M \ times} \ \  c \  \ 0\right]^{T}, \ \ 0 < c \leq KN_tT. \label{STORMMdef1}\\ 
  \Pb&=&\left[P_1\ \ \dots \ \ P_{M} \ \ P_{M+1} \ \ P_{M+2}\right]^{T} \label{STORMMdef2}
   \end{eqnarray}
where $M$ denotes the number of entries in $\db$ that are equal to $KN_tT$. 
Note that when $P_{M+1}=0$, the constellation point corresponding to the entry $c$ such that $0 < c < KN_tT$, is not transmitted. We know that whenever (\ref{STORMhalfplane}) is strict, there cannot be an extreme point $\db$ of the constraint set formed by (\ref{STORMhalfplane}) and (\ref{STORMhypercube}), which has an entry $c$ such that $0 < c < KN_tT$. 
 Therefore, in the case of a strict half-plane constraint, we will take $P_{M+1}=0$ for the optimal constellation without any loss of generality, which   simplifies the subsequent convex minimization problem.   The cardinality of the constellation $L$ depends on the number of non-zero probabilities in the optimal constellation and is related to $M$ by $L \leq M+2$ in general.


   With the structure of the optimal $\db$, the optimal set of probabilities are determined next in terms of $M$ and $c$. Following that, the values of $M$ and $c$ are obtained that maximize the resulting objective function. For convenience, consider minimizing the negative of the objective function in (\ref{STORMsimplifiedmax}) after the optimal $\db$ is substituted as follows:
  \begin{eqnarray}
  & \displaystyle \min_{\{P_i\}_{i=1}^{M+2}} & -K^2N_t^2T^2\sum_{i=1}^{M} P_i(1-P_i)  - c^2 P_{M+1}(1-P_{M+1}). \label{STORMconvex}\\
  &\mbox{subject to}& KN_tT \sum_{i=1}^{M}  P_i + c P_{M+1} \leq E \\
  &  &\sum_{i=1}^{M+2} P_i = 1 , \ \ \ P_i \geq 0 , \ 1 \leq i \leq M+2 
  \end{eqnarray}
  The optimization over $\Pb$  in (\ref{STORMconvex}) is the more commonly studied convex minimization problem \cite{BoydS:Optim}. The Lagrangian can be written as
  \begin{eqnarray}
  L(\Pb,\beta,\lambda,\{\mu_i\}_{i=1}^{M+2})&=& -K^2N_t^2T^2\sum_{i=1}^{M} P_i(1-P_i)  - c^2 P_{M+1}(1-P_{M+1}) +  \beta\left(\sum_{i=1}^{M+2} P_i - 1\right) \nonumber \\
  & &   + \lambda\left\{KN_tT \sum_i^M P_i + c P_{M+1} - E\right\} - \sum_{i=1}^{M+2} \mu_i P_i.
  \end{eqnarray}
It can be verified that Slater's conditions \cite{BoydS:Optim} are satisfied and hence, strong duality holds. Therefore, the Karush-Kuhn-Tucker (KKT) conditions are both necessary and sufficient for the optimal solution $\Pb$ and are given as
  \begin{eqnarray*}
  \lambda \geq 0 \hspace{5mm}, \hspace{5mm} \mu_i \geq 0 \ \ \forall i \hspace{5mm}, \hspace{5mm} KN_tT \sum_{i=1}^{M} P_i + c P_{M+1} &\leq& E\\
  \sum_{i=1}^{M+2} P_i = 1 \hspace{3mm}, \hspace{3mm} \lambda\left\{KN_tT \sum_{i=1}^{M} P_i + c P_{M+1} - E\right\} &=& 0 \hspace{2mm} , \hspace{3mm} \mu_i P_i = 0 \ ,\ 0 \leq i \leq M+2\\
   -K^2N_t^2T^2(1-2P_i) + \lambda KN_tT + \beta - \mu_i &=& 0 \ \ ,  \ \ \ 1 \leq i \leq M\\
   -c^2(1-2P_{M+1}) + \lambda c + \beta - \mu_{M+1}&=& 0 \ \ ,\\
   \beta - \mu_{M+2}&=&0 \ .
  \end{eqnarray*}  
  By eliminating the slack variable $\mub$, we get
  \begin{eqnarray}
   K^2N_t^2T^2(2P_i-1) + \lambda KN_tT + \beta &\geq& 0 \ , \ \ 1 \leq i \leq M \label{STORMone}\\
  c^{2}(2P_{M+1}-1) + \lambda c + \beta &\geq& 0 \label{STORMtwo}\\
  \beta &\geq& 0 \label{STORMthree}\\
  \lambda\left(KN_tT \sum_{i=1}^{M} P_i + c P_{M+1} - E\right)&=&0 \label{STORMfour}\\
  \beta P_{M+2} &=&0 \label{STORMfive}\\
  \left(K^2N_t^2T^2(2P_i-1) + \lambda KN_tT + \beta\right)P_i&=&0 \ , \ \ 1 \leq i \leq M \label{STORMsix}\\
  \left(c^2(2P_{M+1}-1) + \lambda c + \beta\right)P_{M+1}&=&0 \label{STORMseven}\\
  P_i&\geq&0 \ , \ \ 1 \leq i \leq M+2 \label{STORMeight}\\
  \sum_{i=1}^{M+2} P_i &=&1 \label{STORMnine} \\
  KN_tT \sum_{i=1}^{M} P_i + c P_{M+1} &\leq& E \ .\label{STORMten}
  \end{eqnarray}  

From (\ref{STORMsix}), it can be seen that $P_i$ can take one of two values, namely, 
  \begin{eqnarray}
  P_i=0 \qquad {\rm or} \qquad P_i&=&\frac{1}{2} - \frac{\lambda KN_tT + \beta}{2K^2N_t^2T^2} . \label{STORMprobi}
  \end{eqnarray}
Points with zero probability are redundant and since the optimal number $M$ is determined only later, it may be assumed that the $M$ probabilities $P_i$ for $1 \leq i \leq M$ are the same and given in (\ref{STORMprobi}) and denote these probabilities simply as ``$P_i$''. Similarly from (\ref{STORMseven}), $P_{M+1}$ can take one of two values,  namely,
  \begin{eqnarray}
  P_{M+1}=0 \qquad {\rm or} \qquad P_{M+1}&=&\frac{1}{2} - \frac{\lambda c + \beta}{2c^2} . \label{STORMprobc}
  \end{eqnarray}

Four cases must be considered to find the solutions to the KKT conditions. Recall that $K N_t T \geq E$.

  \emph{Case 1 : }  \hspace{3mm} $KN_tT \sum_{i=1}^{M} P_i + c P_{M+1} < E \ \ , \ \ P_{M+2}=0$.

   The strict inequality in (\ref{STORMten}) implies that $\lambda=0$ from (\ref{STORMfour}). Since the  power constraint is a strict inequality, we may  take $P_{M+1}=0$ from the discussion that follows (\ref{STORMMdef2}). Therefore, $P_i=\frac{1}{M}$ is necessary to satisfy (\ref{STORMnine}).   From (\ref{STORMprobi}), we obtain $\beta=\frac{M-2}{M}K^2N_t^2T^2$.  The condition $\beta \geq 0$  implies that $M \geq 2$. The strict inequality in (\ref{STORMten}) together with $P_{M+1}=0$ implies that this case holds when $KN_tT < E$, which is never true. Therefore, this case does not occur.


  \emph{Case 2 : }  \hspace{3mm} $KN_tT \sum_{i=1}^{M} P_i + c P_{M+1} < E \ \ , \ \ P_{M+2}>0$.

  The strict inequality in (\ref{STORMten}) implies that $\lambda=0$ from (\ref{STORMfour}). Since the  power constraint is a strict inequality, we may  take $P_{M+1}=0$ from the discussion that follows (\ref{STORMMdef2}). Since $P_{M+2}>0$, we have $\beta = 0$ from (\ref{STORMfive}). Therefore, $P_i = \frac{1}{2}$ from (\ref{STORMprobi}) and $P_{M+2} =  \frac{1}{2}$. From (\ref{STORMten}), this case applies when $K N_t T < 2 E$ and $M=1$. 

  \emph{Case 3 : }  \hspace{3mm} $KN_tT \sum_{i=1}^{M} P_i + c P_{M+1} = E \ \ , \ \ P_{M+2}>0$.

  Since $P_{M+2}>0$,  we must have $\beta=0$ by (\ref{STORMfive}). There are three sub-cases here, viz., (i) $P_i > 0 \ , P_{M+1} > 0$ \ (ii) $P_i > 0 \ , P_{M+1}=0$ \ and \ (iii) $P_i=0 \ , P_{M+1}>0$. We first consider sub-case (i).  \newline
(i) Using the values $P_i=\frac{1}{2} - \frac{\lambda}{2KN_tT}$ and   $P_{M+1}=\frac{1}{2} - \frac{\lambda }{2c}$ from (\ref{STORMprobi}) and (\ref{STORMprobc}) in the power constraint equality, we can solve for $\lambda$ as $\lambda=\frac{MK N_t T + c -2E}{M+1}$. Substituting this value of $\lambda$ in (\ref{STORMprobi}) and (\ref{STORMprobc}), we obtain 
  \begin{eqnarray}
  P_i&=&\frac{K N_t T-c+2E}{2(M+1)K N_t T} \label{STORMpicase3}\\
  P_{M+1}&=&\frac{M(c-K N_t T)+2E}{2c(M+1)} \label{STORMpccase3} \ .
  \end{eqnarray}
  Using the above probabilities in the objective function $f$ given in (\ref{STORMconvex}), we observe that 
  \begin{eqnarray}
  \frac{d^2f}{dc^2}&=&-\frac{M}{2(M+1)}  \ \leq \ 0 \ ,
  \end{eqnarray}
   which means that $f$ is a concave function over $c$. Since $P_{M+1} \geq 0$, we get from (\ref{STORMprobc}) that $\lambda \leq c$. Therefore, the range of $c$ in this case is given by $\lambda \leq c \leq K N_t T$. Since the optimization of $f$ over $c$ is a concave minimization problem, the minimum is either at $c=\lambda$ or $c=K N_t T$ by Lemma \ref{STORMextreme}. 

 Choosing $c=\lambda$ gives $P_{M+1}=0$ from (\ref{STORMprobc}), $\lambda= K N_t T-\frac{2E}{M}$ and therefore 
\begin{eqnarray}
P_i=\frac{E}{MK N_t T} \ . \label{STORMclambdai}
\end{eqnarray}
 Consequently, from (\ref{STORMnine}) we get that 
\begin{eqnarray}
P_{M+2}=1-\frac{E}{K N_t T} \ \label{STORMclambdac}.
\end{eqnarray}

If $c$ were instead chosen to be $K N_t T$, then from (\ref{STORMpicase3}) and (\ref{STORMpccase3}), $P_i=\frac{E}{(M+1)K N_t T}$, $P_{M+1}=\frac{E}{(M+1)K N_t T}$ and therefore $P_{M+2}=1-\frac{E}{K N_t T}$. Since we are yet to optimize over $M$, the above solution clearly is identical to that obtained in (\ref{STORMclambdai}) and (\ref{STORMclambdac}). So we may choose $c=\lambda$ itself as the solution.

 For $c=\lambda$,  since $\lambda \geq 0$, this case requires $K N_t T \geq \frac{2E}{M}$. Moreover, the power constraint equality requires that  $K N_t T \geq E$. Hence, this sub-case solves the convex optimization problem for the cases $K N_t T \geq E \ , \ M \geq 2$ and $K N_t T \geq E \ , \ M = 1$. 


 Even for sub-cases (ii) and (iii), it can be easily verified that we get essentially the same solutions as the previous sub-case.
  
  \emph{Case 4 : }  \hspace{3mm} $KN_tT \sum_{i=1}^{M} P_i + c P_{M+1} = E \ \ , \ \ P_{M+2}=0$. 


The cases $K N_t T \geq E \ , \ M \geq 2$ and $K N_t T \geq E \ , \ M = 1$  are solved completely through Cases 2 and 3. This is true because by strong duality, the constellations obtained in Cases 2 and 3 are both necessary and sufficient for optimality. Moreover, since $K N_t T < E$ does not occur, we do not solve for Case 4 since we will get no new solutions or insights. 



  The last step is to find the best possible $M$. We revert to the problem which is a maximization of the objective function $f$ for convenience. From Case 3, which yields the only pertinent solution for $T\geq 2$,   the  objective function with the optimal probabilities given in (\ref{STORMclambdai}) and (\ref{STORMclambdac}) is
  \begin{eqnarray}
  f&=&KN_tTE\left(1-\frac{E}{MKN_tT}\right) \ . \label{STORMcardinalitylargepeak}
  \end{eqnarray}

Notice that $f$ is an increasing function of $M$, and $M$ needs to be chosen as large as possible. However, if $M$ is chosen so that $M>T$, inequality (\ref{STORMnegzero}) would be strict since it is not possible to make the columns of all pairs of different constellation matrices orthogonal. Therefore, $M=T$ is optimal among $M$ satisfying $M \leq T$. When we take the limit as $M \to \infty$, we get an upper bound on the mutual information which is not achievable (hence the strict inequality for the upper bound in (\ref{STORMMIlargepeak})).

  To complete the proof, notice that we may use the jointly optimal $\Pb$ and $\db$ with the structure of constellation points given in (\ref{STORMdftconst1},\ref{STORMdftconst2}) so that the upper bound in  (\ref{STORMnegzero}) is achieved with equality when $M \leq T$. Therefore, the optimal constellations have been obtained for the case $M \leq T$. When $M \geq T$, we can obtain an upper bound on the maximum achievable mutual information by letting $M \to \infty$ in (\ref{STORMcardinalitylargepeak})  (and multiplying by the factor $\frac{N_r}{2T}$).
  \endproof

\subsection{Spectral Efficiency}
Consider the normalized energy per bit for reliable communications which is given as
\begin{eqnarray}
\frac{E_b}{N_0} = \frac{P}{C(P)} \ , \label{STORMenergybit}
\end{eqnarray}
where $C(P)$ is the Shannon capacity for the channel in bits per dimension. For the case when $C(P)$ is a non-decreasing concave function, it can be seen that  (\ref{STORMenergybit}) achieves its minimum value over all $P$, as $P \to 0$. However, this is not true in the PPAPR-constrained case. Indeed, since the capacity is $O(P^2)$, $\frac{E_b}{N_0} \to \infty$ as $P \to 0$. Therefore, it is not energy-efficient to operate at asymptotically low SNR in this case. The mutual information of STORM at any SNR is
\begin{eqnarray}
I^{STORM}(P)&=&\sum_i \ P_i \  \E_{\Yb | \Xb_i}\left[\log\left(\frac{p(\Yb | \Xb_i)}{\sum_j P_j  \ p(\Yb | \Xb_j)}\right)\right]. \label{STORMMonte-Carlo}
\end{eqnarray}
The expectations in (\ref{STORMMonte-Carlo}) can be calculated using Monte-Carlo integration. Thus the normalized energy per bit required for STORM can be determined as $\frac{E_b}{N_0} = \frac{P}{I^{STORM}(P)}$, over the entire range of SNRs. It can be seen through extensive simulations over a variety of cases that the minimum energy per bit typically occurs at a low but non-vanishing SNR. STORM should hence be used in the vicinity of this SNR, for maximum spectral efficiency. In the absence of the capacity of the noncoherent MIMO channel at a general SNR however, there is no fair yard stick to compare the energy per bit of STORM against that of the capacity achieving scheme.

 \section{The peak-constrained case} \label{STORMCUE}
In this section, the \emph{peak-constrained} problem is considered where the peak constraint $K$ in (\ref{STORMpeak-power}) is a fixed constant, {\em independent} of the average power $P$. It can be shown by a simple time-sharing argument that the channel capacity in this case is concave and non-decreasing in $P$. 
Therefore, the normalized energy per bit $\frac{E_b}{N_0}$ given in (\ref{STORMenergybit}) can be seen to attain its minimum value over all $P$, as $P \to 0$. Let us denote the normalized minimum energy per bit for our channel model by $\frac{E_b}{N_0}_{min}$, in keeping with common usage \cite{VerduS:efficiency}. Since $C(P)$ is a non-decreasing function of $P$,  it can be assumed without any loss of generality that the average power constraint is $\frac{1}{T}\E\left[\tr(\Xb\Xb^{*})\right] = P$ instead of $\frac{1}{T}\E\left[\tr(\Xb\Xb^{*})\right] \leq P$. 
The capacity function (in bits/dimension) admits the following Taylor series expansion
\begin{eqnarray}
C(P) = \dot{C}(0) P \log_2 e + \frac{1}{2} \ddot{C}(0) P^2 \log_2 e + o(P^2) \ , \label{STORMgencap}
\end{eqnarray}
where $\dot{C}(0)$ and $\ddot{C}(0)$ are the first and second derivatives of $C(P)$ computed in nats/dimension. The notation and units introduced above for $C(P)$, $\dot{C}(0)$ and $\ddot{C}(0)$ will be used in the rest of this paper. The  capacity per unit energy (in bits per joule)  is the reciprocal of ${\frac{E_b}{N_0}}_{min}$, and is equal to $\dot{C}(0)\log_2 e$ in the peak-constrained case, and either metric can be considered to be a measure of energy efficiency. Therefore, the notions of minimizing the energy per bit and maximizing the information rate per unit energy will be used interchangeably. The minimization of energy per bit is considered in Section \ref{sec-maxinforate}. Note however that since this minimum occurs at a vanishing SNR, a fixed rate (in bits/sec) of communication can be only achieved in the limit of infinite bandwidth. It is hence of interest to communicate at low but non-vanishing SNR and also do so in a bandwidth efficient manner, which brings us to the notion of wideband slope introduced in \cite{VerduS:efficiency}.

The slope of the capacity function versus $\frac{E_b}{N_0}$ (also called the spectral efficiency function) in bits per second per hertz per 3 dB at zero spectral efficiency is defined as the wideband slope in \cite{VerduS:efficiency} and was shown to be given in terms of $\dot{C}(0)$ and $\ddot{C}(0)$ as
\begin{eqnarray}
{\mathcal{S}}_0 = \frac{2\left[\dot{C}(0)\right]^2}{-\ddot{C}(0)} \ .
\end{eqnarray}
The motivation for considering the wideband slope as a performance metric is that, 
while achieving $\frac{E_b}{N_0}_{min}$ is desirable for energy efficiency, the rate of convergence of $\frac{E_b}{N_0}$ to $\frac{E_b}{N_0}_{min}$ as $P \to 0$ is also an important factor at low $P$, which in turn is closely tied to spectral efficiency. 
The higher the wideband slope, the greater is the spectral efficiency when operating at small but non-vanishing SNR.
This point about the importance of the wideband slope was highlighted through several examples in the insightful work of \cite{VerduS:efficiency}. An important example provided there was that of noncoherent communications with an input average power constraint alone, and the wideband slope in this case was found to be ${\mathcal{S}}_0 = 0$ in contrast to that of coherent communication where it is positive. This result implies that to approach $\frac{E_b}{N_0}_{min}$, the bandwidth for reliable noncoherent communications becomes prohibitively large and the associated signaling scheme prohibitively peaky, and therefore not realistic (i.e., bandwidth limited and peak-limited) scheme can achieve $\frac{E_b}{N_0}_{min}$. 

In this work, the noncoherent MIMO channel is considered with a peak-constraint on the input, in addition to the average power constraint. It is shown that with the additional peak-constraint, which is necessary for meaningful results at low SNR, there is a tradeoff between the minimum energy per bit and the wideband slope. This provides a far more detailed characterization of the wideband slope than if only the average power constraint were imposed, and in particular it shows that it is possible to have ${\mathcal{S}}_0 > 0$  provided the peak-constraint on the input is less than a certain constant. In the process, the $T+1$ point constellation is derived in Section \ref{sec-optwdbnslp} from among constellations that achieve minimum energy per bit (or equivalently, $\dot{C}(0)$) that is {\em optimal} in wideband slope (or maximize $\ddot{C}(0)$), which interestingly, turns out to be STORM again. STORM is hence optimal in spectral efficiency in the wideband regime. Apart from providing fundamental limits on peak-limited MIMO noncoherent communications, our results and conclusions also temper the pessimistic conclusions that result from the consideration of noncoherent communication under just an average power constraint \cite{VerduS:efficiency}. 

\subsection{Achieving minimum energy per bit} \label{sec-maxinforate}

In this section, the necessary and sufficient conditions for a constellation to achieve  $\frac{E_b}{N_0}_{min}$ are derived. First, the following definition and lemma are needed from optimization theory \cite{BoydS:Optim , HorstR:glo}.
\begin{definition} \label{STORMdefnquasi}
A function $f$ is strictly quasiconcave over a convex set $\Ak$ iff for any $\xb_1 , \xb_2 \in \Ak$,  and for $0 < \theta < 1$,
\begin{eqnarray}
f(\theta \xb_1 + (1-\theta) \xb_2) > \min\left\{f(\xb_1) , f(\xb_2)\right\} \ . \label{STORMquasiconcave}
\end{eqnarray}  
\end{definition} 
  \begin{lemma}
  The global minimum of a strictly quasiconcave function $f$ over a compact convex set $\Ak$ is attained at a point $\xb \in \Ak$ only if $\xb$ is an extreme point of $\Ak$.  \label{STORMquasiextreme}
  \end{lemma}
\begin{theorem} \label{STORMnecsuffPL}
 Consider a constellation $\Ck$ with non-zero matrices $\{\Xb_i\}_{i=1}^{L-1}$ and respective probabilities $\{P_i\}_{i=1}^{L-1}$, and the zero matrix with probability $P_0$. Let $\Ck$ satisfy the average power constraint $\E\left[\tr(\Xb\Xb^{*})\right] = PT = E$ and the peak-constraint (\ref{STORMpeak-power}) as in the peak-constrained problem.  Then, $\Ck$ achieves  the capacity per unit energy as $P \to 0$ iff  its constellation matrices and respective probabilities are of the following form
 \begin{eqnarray}
 \Xb_i &=& \sqrt{K} \ \vb_i \wb_i^{*}  \ \ , \ \ 1 \leq i \leq L-1 \label{STORMppnonzeropoints} \\
 \Xb_0 &=& \Zrb_{T \times N_t}  \ , \label{STORMpplargepeakopt} \\
\sum_{i=1}^{L-1} P_i &=& \frac{P}{KN_t} \\
P_0 &=& 1 - \frac{P}{KN_t} \ ,
 \end{eqnarray}
  where for each $i$, $\vb_i \in \Cplx^{T \times 1}$, $\wb_i \in \Cplx^{N_t \times 1}$ and $|\left[\vb_i \wb_i^{*}\right]_{mn}| = 1 \ \forall \ i, m, n$.
The  capacity per unit energy achieved by the above constellation is 
\begin{eqnarray}
 N_r \cdot \left\{ 1 - \frac{\log (1 + K N_tT)}{K N_tT}\right\} \log_2 e  \ \ \ \ \mbox{bits/joule} \ . \label{STORMcapune}
\end{eqnarray}
\end{theorem} 
\proof
Let the mutual information  between $\Ck$ and the output $\Yb$ be denoted as $I(P)$ (in nats per dimension).  It is known from \cite{VerduS:unitcost} that to achieve the capacity per unit energy, it is sufficient to use one symbol apart from the zero energy symbol. Therefore, our formulation, which assumes a discrete input with an arbitrary number of points, is without any loss of generality. The optimization problem that is to be solved is given as
\begin{eqnarray}
  &\max_{\{P_i\}_{i=0}^{L-1}, \{\Xb_i\}_{i=1}^{L-1}}& \ \dot{I}(0)  \ , \label{STORMpeaklimEb1}\\
   &\mbox{subject to}&   \sum_{i=0}^{L-1} P_i \ \tr(\Xb_i\Xb_i^{*}) = PT \ , \  \|\Xb_i\|_{\infty} \leq \sqrt{K}, \ \forall i \nonumber\\
  && \ \sum_{i=1}^{L-1} P_i = 1 - P_0, \ P_i \geq 0  \ \ \forall i \nonumber \ .
  \end{eqnarray}
A general formula for $\dot{I}(0)$ was derived in \cite{VerduS:efficiency} and is given as
\begin{eqnarray}
\dot{I}(0) = \lim_{P \to 0} \frac{\E_{\Xb} \left[D(P_{\Yb | \Xb} || P_{\Yb | \Xb=\Zrb})\right]}{\E_{\Xb} \left[\tr(\Xb \Xb^{*})\right]} \ .
\end{eqnarray}
Since 
\begin{eqnarray}
\displaystyle \max_{\{P_i\}_{i=0}^{L-1}, \{\Xb_i\}_{i=1}^{L-1}} \ \lim_{P \to 0} \frac{\E_{\Xb} \left[D(P_{\Yb | \Xb} || P_{\Yb | \Xb=\Zrb})\right]}{\E_{\Xb} \left[\tr(\Xb \Xb^{*})\right]} \leq \lim_{P \to 0} \ \max_{\{P_i\}_{i=0}^{L-1}, \{\Xb_i\}_{i=1}^{L-1}}  \frac{\E_{\Xb} \left[D(P_{\Yb | \Xb} || P_{\Yb | \Xb=\Zrb})\right]}{\E_{\Xb} \left[\tr(\Xb \Xb^{*})\right]} \ ,
\end{eqnarray}
an upper bound for the optimal value of the problem in (\ref{STORMpeaklimEb1}) is
\begin{eqnarray}
 \lim_{P \to 0}  \hspace{5mm} &\max_{\{P_i\}_{i=0}^{L-1}, \{\Xb_i\}_{i=1}^{L-1}}& \ \frac{\E_{\Xb}\left[D(P_{\Yb | \Xb} || P_{\Yb | \Xb=\Zrb})\right]}{\E_{\Xb} \left[\tr(\Xb \Xb^{*})\right]} \ , \label{STORMpeaklimEb2}\\
   &\mbox{subject to}&    \sum_{i=0}^{L-1} P_i \ \tr(\Xb_i\Xb_i^{*}) = PT \ , \ \|\Xb_i\|_{\infty} \leq \sqrt{K}, \ \forall i \nonumber\\
  && \ \sum_{i=1}^{L-1} P_i = 1 - P_0, \ P_i \geq 0  \ \ \forall i \nonumber \ .
  \end{eqnarray}
The objective function in (\ref{STORMpeaklimEb2}) can be evaluated as
\begin{eqnarray}
 \frac{\E_{\Xb}\left[D(P_{\Yb | \Xb} || P_{\Yb | \Xb=\Zrb})\right]}{\E_{\Xb} \left[\tr(\Xb \Xb^{*})\right]} &=& N_r \cdot \left\{ 1 - \frac{\sum_{i=1}^{L-1} P_i \log \det \left(\Ib + \Xb_i\Xb_i^{*}\right)}{\sum_{i=1}^{L-1} P_i \ \tr(\Xb_i\Xb_i^{*})}\right\} \ .
\end{eqnarray}
Consequently, the problem that needs to be solve is
  \begin{eqnarray}
  &\min_{\{P_i\}_{i=0}^{L-1}, \{\Xb_i\}_{i=1}^{L-1}}& \  \frac{\sum_{i=1}^{L-1} P_i \log \det \left(\Ib + \Xb_i\Xb_i^{*}\right)}{\sum_{i=1}^{L-1} P_i \ \tr(\Xb_i\Xb_i^{*})} \ , \label{STORMinterebn00}\\
   &\mbox{subject to}& \ \sum_{i=0}^{L-1} P_i \ \tr(\Xb_i\Xb_i^{*}) = PT  \ , \ \|\Xb_i\|_{\infty} \leq \sqrt{K}, \ \forall i \nonumber\\
  && \ \sum_{i=1}^{L-1} P_i = 1 - P_0, \ P_i \geq 0  \ \ \forall i \nonumber \ .
  \end{eqnarray} 
Relaxing the peak constraint, the optimal value of the problem in (\ref{STORMinterebn00}) over the signal constellation (but with the probabilities fixed) is lower bounded by the optimal value of the problem
  \begin{eqnarray}
  &\min_{ \{\Xb_i\}_{i=1}^{L-1} , \{d_i\}_{i=1}^{L-1} , \{P_i\}_{i=0}^{L-1}}& \  \frac{\sum_{i=1}^{L-1} P_i \log \det \left(\Ib + \Xb_i\Xb_i^{*}\right)}{\sum_{i=1}^{L-1} P_i \  d_i} \ ,\label{STORMinterebn01}\\
   &\mbox{subject to}&   \sum_{i=0}^{L-1} P_i \ d_i = PT  \ , \ \tr(\Xb_i\Xb_i^{*}) = d_i \ , \  0 \leq d_i \leq K N_t T, \ \forall i \ .\nonumber
  \end{eqnarray}
The optimal values of problems (\ref{STORMinterebn00}) and (\ref{STORMinterebn01}) are the same iff the $\{\Xb_i\}_{i=1}^{L-1}$ that solves (\ref{STORMinterebn01}) also satisfies $\|\Xb_i\|_{\infty} \leq \sqrt{K} \ , \ \forall i$. 

As in the PPAPR constrained problem, the above problem can be solved as a two-stage optimization, where in the first stage, the probabilities $\{P_i\}_{i=0}^{L-1}$ are fixed and the constellation $\{\Xb_i\}_{i=1}^{L-1}$ is optimized. In the second step, the resulting objective function over is optimized over $\{P_i\}_{i=0}^{L-1}$. 


Consider a fixed, feasible but otherwise arbitrary $\{P_i\}_{i=0}^{L-1}$. It can be verified that for each $i$, 
\begin{eqnarray}
\min_{\tr(\Xb_i\Xb_i^{*})=d_i} \log \det \left(\Ib+\Xb_i\Xb_i^{*}\right) = \log(1+d_i) \ ,
\end{eqnarray}
is solved iff $\Xb_i$ has unit rank. 

Therefore, the problem in (\ref{STORMinterebn01}) can be re-written as
   \begin{eqnarray}
  &\min_{ \{d_i\}_{i=1}^{L-1} , \{P_i\}_{i=0}^{L-1}}& \  \frac{\sum_{i=1}^{L-1} P_i \log (1+d_i)}{\sum_{i=1}^{L-1} P_i \  d_i} \ ,\label{STORMinterebn02}\\
   &\mbox{subject to}&   \sum_{i=1}^{L-1} P_i \ d_i =  PT , \  0 \leq d_i \leq K N_t T, \ \forall i \nonumber
  \end{eqnarray}
Let $\db = [d_1 \  d_2 \ \dots \ d_{L-1}]^{T}$.
Consider the set 
\begin{eqnarray}
\Ak_t  = \left\{\db : h(\db) = \frac{\sum_{i=1}^{L-1} P_i \log (1+d_i)}{\sum_{i=1}^{L-1} P_i \  d_i} > t \ , \ d_i \geq 0 \ \forall i , \ t \geq 0\right\} \ .
\end{eqnarray}
 Since $\sum_{i=1}^{L-1} P_i \log (1+d_i) - t \sum_{i=1}^{L-1} P_i \  d_i$ is strictly concave for every real $t$, the set $\Ak_t$ is  convex. Therefore, considering any two points $\db_1, \db_2 \in \Ab_t$ where $t = \min \left\{h(\db_1) , h(\db_2)\right\}$ and using  Definition \ref{STORMdefnquasi},  $\frac{\sum_{i=1}^{L-1} P_i \log (1+d_i)}{\sum_{i=1}^{L-1} P_i \  d_i}$ is a strictly quasiconcave function of $\db$. Hence, from Lemma \ref{STORMquasiextreme}, the solution of (\ref{STORMinterebn02}) is achieved at a vertex  of the constraint set. Using Lemma \ref{STORMvertexchar}, each vertex of the constraint set consists of $L-1$ entries that are either $K N_tT$ or $0$, and exactly  one entry $c$ such that  $0 \leq c \leq KN_tT$. 




It can therefore be assumed, without loss of generality, that the optimal $\db$ and the corresponding probabilities are
  \begin{eqnarray}
  \db&=& \left[\underbrace{KN_tT \  \ \dots \  \ KN_tT}_{M \ times} \ \  c \  \ 0\right]^{T}, \ \ 0 \leq c \leq KN_tT. \label{STORMMdef1wb}\\ 
  \Pb&=&\left[P_1\ \ \dots \ \ P_{M} \ \ P_{c} \ \ P_{0}\right]^{T}. \label{STORMMdef2eb}
   \end{eqnarray}
where, for convenience, the symbol $M$ is introduced to denote the number of entries in $\db$ that are equal to $KN_tT$. Since the objective function is a symmetric function of $\db$, the specific arrangement of the entries is immaterial. Using this structure for $\db$, the problem in (\ref{STORMinterebn02}) can be re-written and bounded from below as
    \begin{eqnarray}
  &\displaystyle \min_{c, \{P_i\}_{i=0}^{L-1}}& \  \frac{\sum_{i=1}^{M} P_i \log (1+KN_tT) + P_c \log (1+c) }{\sum_{i=1}^{M} P_i  \  KN_tT +  c \ P_c} \ ,\\
   &\mbox{subject to}&    \ 0 \leq c \leq KN_tT \ , \ \sum_{i=1}^{M} P_i KN_tT + c\  P_c \  =  PT \ , \ \sum_{i=1}^{M} P_i = 1 - P_0 \nonumber \\ \label{STORMinterebn03}
\geq &\displaystyle \min_{c, \{P_i\}_{i=0}^{L-1} \atop 0 \leq c \leq KN_tT}&  \  \frac{\sum_{i=1}^{M} P_i \log (1+KN_tT) + P_c \log (1+c) }{\sum_{i=1}^{M} P_i  \  KN_tT +  c \ P_c} \ . 
  \end{eqnarray} 
The problem in (\ref{STORMinterebn03}) is easily seen to be the minimization of a strictly quasiconcave function over $c$. Therefore, the solution has to be among the vertices of $0 \leq c \leq KN_tT$, ie., either $c=0$ or $c=KN_tT$. Notice that with either choice of $c$, the objective function is $\frac{\log(1+KN_tT)}{KN_tT}$, and is independent of $\{P_i\}_{i=0}^{L-1}$. Therefore, the upper bound on the optimal value of the problem in (\ref{STORMpeaklimEb1}) is
\begin{eqnarray}
N_r \cdot \left\{ 1 - \frac{\log (1 + K N_tT)}{K N_tT}\right\} \ . \label{STORMPLUB}
\end{eqnarray}
Since $d_i = KN_tT \ \ \forall i$, for equality to hold in the inequality leading to (\ref{STORMinterebn01}), it is necessary and sufficient that the non-zero matrices  $\{\Xb_i\}_{i=1}^{L-1}$ be of the form
  \begin{eqnarray}
    \Xb_i&=&\sqrt{K} \ \vb_i \wb_i^{*}   \ \ \forall i, \label{STORMrankoneX}
    \end{eqnarray}
 where $\vb_i \in \Cplx^{T \times 1}$,  $\wb_i \in \Cplx^{N_t \times 1}$ are such that  $|\left[\vb_i \wb_i^{*}\right]_{mn}| = 1 \ \forall \ i, m, n$.
By substituting  (\ref{STORMrankoneX}) in (\ref{STORMpeaklimEb1}), a lower bound on the optimal value of (\ref{STORMpeaklimEb1}) is obtained, which coincides with the upper bound in (\ref{STORMPLUB}), implying that (\ref{STORMPLUB}) is the optimal value of the problem in (\ref{STORMpeaklimEb1}).  From the power constraint, $\sum_{i=1}^{L-1} P_i = \frac{P}{KN_t}$ must be true and $P_0 = 1 - \frac{P}{KN_t} > 0$.  Therefore, it can be  concluded that
\begin{eqnarray}
\dot{C}(0) = N_r \cdot \left\{ 1 - \frac{\log (1 + K N_tT)}{K N_tT}\right\}   \ .
\end{eqnarray}
\endproof
Note that the capacity per unit energy in (\ref{STORMcapune}) is independent of the number of points $L$. In particular, it can be achieved with a 2-point constellation.

{\emph{Corollary :}}
The following two point constellation achieves the capacity per unit energy as the average power $P \to 0$
 \begin{eqnarray}
 \left(\Xb_1 , P_1\right)&=& \left(\sqrt{K} \ \vb \  \wb^{*}  \ , \ \frac{P}{KN_t}\right) \label{STORMMIMOOOKnonzero1}\\
 \left(\Xb_2 , P_2\right)&=& \left(\Zrb_{T \times N_t} \ , \ 1 - \frac{P}{KN_t}\right) \ , \label{STORMMIMOOOKzero1}
 \end{eqnarray} 
where $\vb$ and $\wb$ are column vectors such that $|\left[\vb \wb^{*}\right]_{mn}| = 1 \ \forall \ i, m, n$. \endproof

The above 2-point constellation is referred to as MIMO-OOK (on-off keying).  This constellation can also be obtained directly through a simplified general formula for the capacity per unit energy  derived in \cite{VerduS:unitcost}. It turns out that the simplified formula in \cite{VerduS:unitcost} can  be evaluated using similar techniques to those used in the proof of Theorem \ref{STORMnecsuffPL}, and is also a more direct approach than the derivation of the capacity per unit energy in \cite{WuX:MIMOexp}. For the sake of completeness, it is given in Appendix\ref{STORMmimoook}.


Clearly, Theorem \ref{STORMnecsuffPL} implies that there is a large class of constellations which achieve  $\frac{E_b}{N_0}_{min}$. For instance, the cardinality can be any $L \geq 2$. Moreover, only the sum of probabilities of the non-zero points is constrained to be $\frac{P}{K N_t}$, while the individual  probabilities can be arbitrary. Further, there is no restriction on the relationship between $\Xb_i$ and $\Xb_j, \; \forall j \neq i$. In particular, $\Xb_i $ can be taken to be all equal to a unit rank matrix $\Xb $ with elements of equal magnitude (equal to $\sqrt{K}$) for all $i =1, 2, \dots, L-1$. In this case, the non-zero points would coincide and become one non-zero point with probability $\frac{P}{K N_t}$, thereby reducing to the 2-point MIMO-OOK constellation of Corollary 3.

\subsection{Maximizing the wideband slope} \label{sec-optwdbnslp}

A key insight provided by \cite{VerduS:efficiency} is that even though different schemes may achieve ${\frac{E_b}{N_0}}_{min}$, an analysis of their wideband slopes could reveal vast differences in the rate of growth of their energy efficiencies around ${\frac{E_b}{N_0}}_{min}$, and therefore differentiates their spectral efficiencies. The wideband slope, which is the measure of spectral efficiency at low but non-vanishing SNR, is therefore critical in the analysis of wideband channels.  Our next aim is therefore, to optimize the wideband slope over constellations which achieve $\frac{E_b}{N_0}_{min}$. 
The next theorem provides a formula for the wideband slope ${\mathcal{S}}_0$ when evaluated for an arbitrary generalized OOK constellation.
\begin{theorem} \label{STORMproppearson}
 Consider a constellation $\Ck$ with non-zero matrices $\{\Xb_i\}_{i=1}^{M}$ and respective probabilities $\{P_i\}_{i=1}^{M}$, and the zero matrix with probability $P_0$. Then
 \begin{equation} 
  {\mathcal{S}}_0= \left\{ 
  \begin{array}{l}
  \frac{2}{T} \frac{N_r^2 \left(\sum_{i=1}^{M} P_i \tr(\Xb_i\Xb_i^{*}) - \sum_{i=1}^{M} P_i \log \det \left(\Ib + \Xb_i\Xb_i^{*}\right) \right)^2}{\sum_{i=1}^M \frac{P_i^2}{(1-P_0)^2} \frac{1}{\left|\Ib - \Xb_i\Xb_i^{*}\Xb_i\Xb_i^{*}\right|^{N_r}} + \sum_i \sum_{j\neq i} \frac{P_i P_j}{(1-P_0)^2} \frac{1}{\left|\Ib - \Xb_i\Xb_i^{*}\Xb_j\Xb_j^{*}\right|^{N_r}} -1} \ , \\
  {\rm if} \; \Ib - \Xb_i\Xb_i^{*}\Xb_j\Xb_j^{*} \; {\rm is~positive~definite} \ \forall \ i,j \\
  0, \; {\rm otherwise.}
  \end{array} \right.
  \label{STORMs0gen}
 \end{equation}
\end{theorem}

\proof See Appendix \ref{STORMproofproppearson}.

The following corollary indicates a fundamental limitation in approaching the capacity per unity energy for a constellation of arbitrary cardinality.
\begin{corollary} \label{STORMCorollaryWBSlope}
Consider a constellation $\Ck$ with non-zero matrices $\{\Xb_i\}_{i=1}^{M}$ and respective probabilities $\{P_i\}_{i=1}^{M}$, and the zero matrix with probability $P_0$. Let $\Ck$ satisfy the average and peak power constraints in the statement of Theorem \ref{STORMnecsuffPL}. Suppose $\Ck$ achieves the capacity per unit energy. Then the wideband slope ${\mathcal S}_0$  is $0$ when $KN_tT > 1$.
\end{corollary}
\proof Since $\Ck$ achieves the capacity per unit energy, it satisfies the necessary conditions stated in Theorem \ref{STORMnecsuffPL}. From Theorem \ref{STORMproppearson}, the wideband slope is non-zero only when the matrix
\begin{eqnarray}
\Ib-\Xb_i\Xb_i^{*}\Xb_j\Xb_j^{*} \label{STORMposdefwb}
\end{eqnarray}
 is positive definite for all pairs $i,j$. The proof of the corollary follows when the necessary conditions for achieving the capacity per unit energy in Theorem \ref{STORMnecsuffPL} are substituted in  (\ref{STORMposdefwb}) and simplified.  
\endproof




\begin{theorem} \label{STORMmaxwbslope}
Among all constellations of Theorem \ref{STORMnecsuffPL} which achieve $\frac{E_b}{N_0}_{min}$, with $T+1$ points, STORM has the maximum wideband slope. 
\end{theorem}
\proof
Since the constellations under consideration achieve $\frac{E_b}{N_0}_{min}$, the numerator in (\ref{STORMs0gen}) is a fixed constant. Further, given the necessary conditions for the constellation to achieve $\frac{E_b}{N_0}_{min}$, the denominator of the wideband slope can be simplified as
\begin{eqnarray}
\sum_{i=1}^{M} \frac{P_i^2}{(1-P_0)^2} \frac{1}{\left(1- K^2N_t^2T^2\right)^{N_r}}  +  \sum_i \sum_{j\neq i} \frac{P_i P_j}{(1-P_0)^2} \frac{1}{\left|\Ib - \Xb_i\Xb_i^{*}\Xb_j\Xb_j^{*}\right|^{N_r}} -1 \ , \label{STORMdenominatiors0}
\end{eqnarray} 
where the matrices $\{\Xb_i\}_{i=1}^{M}$ are of unit rank with entries of equal magnitude $\sqrt{K}$, and $KN_tT < 1$ (to ensure that $\Ib - \Xb_i\Xb_i^{*}\Xb_j\Xb_j^{*}$ is positive semidefinite $ \forall i, j$). Clearly,  (\ref{STORMdenominatiors0}) is minimized when there exist rank-one matrices $\{\Xb_i\}_{i=1}^{M}$  such that $\Xb_j^{*}\Xb_i = \Zrb \ \forall i, j\neq i$. Such a set exists for $M \leq T$, and is denoted by $\Xb_i = \sqrt{K} \vb_i \wb_i^{*} \ \forall i$, where the definitions for $\vb_i$ and $\wb_i$ are the same as in Theorem \ref{STORMmaxMIlowlarge}. 
The problem that needs to be solved is thus
\begin{eqnarray}
\min_{\{P_i\}_{i=1}^{M} \atop \sum_{i=1}^{M} P_i= 1 - P_0 \ , \  \sum_{i=1}^{M} P_i = \frac{P}{KN_t}} \ \sum_{i=1}^{M} \frac{P_i^2}{(1-P_0)^2} \left\{\frac{1}{\left(1- K^2N_t^2T^2\right)^{N_r}} - 1\right\} \ . \label{STORMwbslopestep}
\end{eqnarray}
The objective function in (\ref{STORMwbslopestep}) can be easily shown to be a Schur-convex function \cite{MarshallAW:ineq} of $\left[ P_1 \ P_2 \ \dots \ P_M\right]$. Hence, the minimum occurs when each of the probabilities $\{P_i\}_{i=1}^{M}$ is equal to  $\frac{1 - P_0}{M}$. The optimal value of (\ref{STORMwbslopestep}) is therefore
\begin{eqnarray}
\frac{1}{M} \left\{\frac{1}{\left(1- K^2N_t^2T^2\right)^{N_r}} - 1\right\} \ . \label{STORMoptwbsl}
\end{eqnarray}
Clearly, $M$ has to be made as large as possible, but to ensure achievablity of the optimal value in  (\ref{STORMwbslopestep}), it can be no greater than $T+1$. Therefore, set $M=T+1$. Evidently, the solution to (\ref{STORMwbslopestep}) when $M > T+1$ would provide an upper bound on the maximum wideband slope.
\endproof
Theorem  \ref{STORMmaxwbslope}   establishes the optimality of STORM among $T+1$ point constellations in the peak-constrained case. This means that STORM is spectrally most efficient among all $T+1$ (or fewer) point constellations that achieve maximum capacity per unit energy in the low SNR regime.  

The following corollary provides the wideband slopes of MIMO-OOK and STORM.
\begin{corollary} \label{STORMSTORMEbmin}
The wideband slopes of MIMO-OOK and STORM are respectively,
\begin{eqnarray}
{\mathcal S}_0^{OOK} = \left\{ \begin{array}{ll}
\frac{2}{T} \frac{N_r^2\left(KN_tT - \log(1+KN_tT)\right)^2}{\frac{1}{\left(1 - K^2N_t^2T^2\right)^{N_r}} - 1} & \mbox{if $KN_tT < 1$} \ ; \\
0 & \mbox{if $KN_tT \geq 1$}\ . \end{array} \right. \label{STORMOOKS0} \\
{\mathcal S}_0^{STORM} =  \left\{ \begin{array}{ll}
2  \frac{N_r^2\left(KN_tT - \log(1+KN_tT)\right)^2}{\frac{1}{\left(1 - K^2N_t^2T^2\right)^{N_r}} - 1} & \mbox{if $KN_tT < 1$} \ ;\\
0 & \mbox{if $KN_tT \geq 1$}\ . \end{array} \right. \label{STORMSTORMS0}
\end{eqnarray}
\end{corollary}
\proof  The wideband slopes follow by substituting the MIMO-OOK and STORM constellations in the result of Theorem \ref{STORMproppearson}.  \endproof

\subsection{Remarks} \label{sec-remwdbndslp}
Since STORM was obtained as the optimal constellation even in the PPAPR constrained case, many of the remarks on STORM following Theorem \ref{STORMmaxMIlowlarge} and in Section \ref{Remarks1} apply even to the peak-constrained case. Here we only state new insights pertinent to the peak-constrained case.

\begin{figure}
\centerline{\scalebox{0.65}{\includegraphics{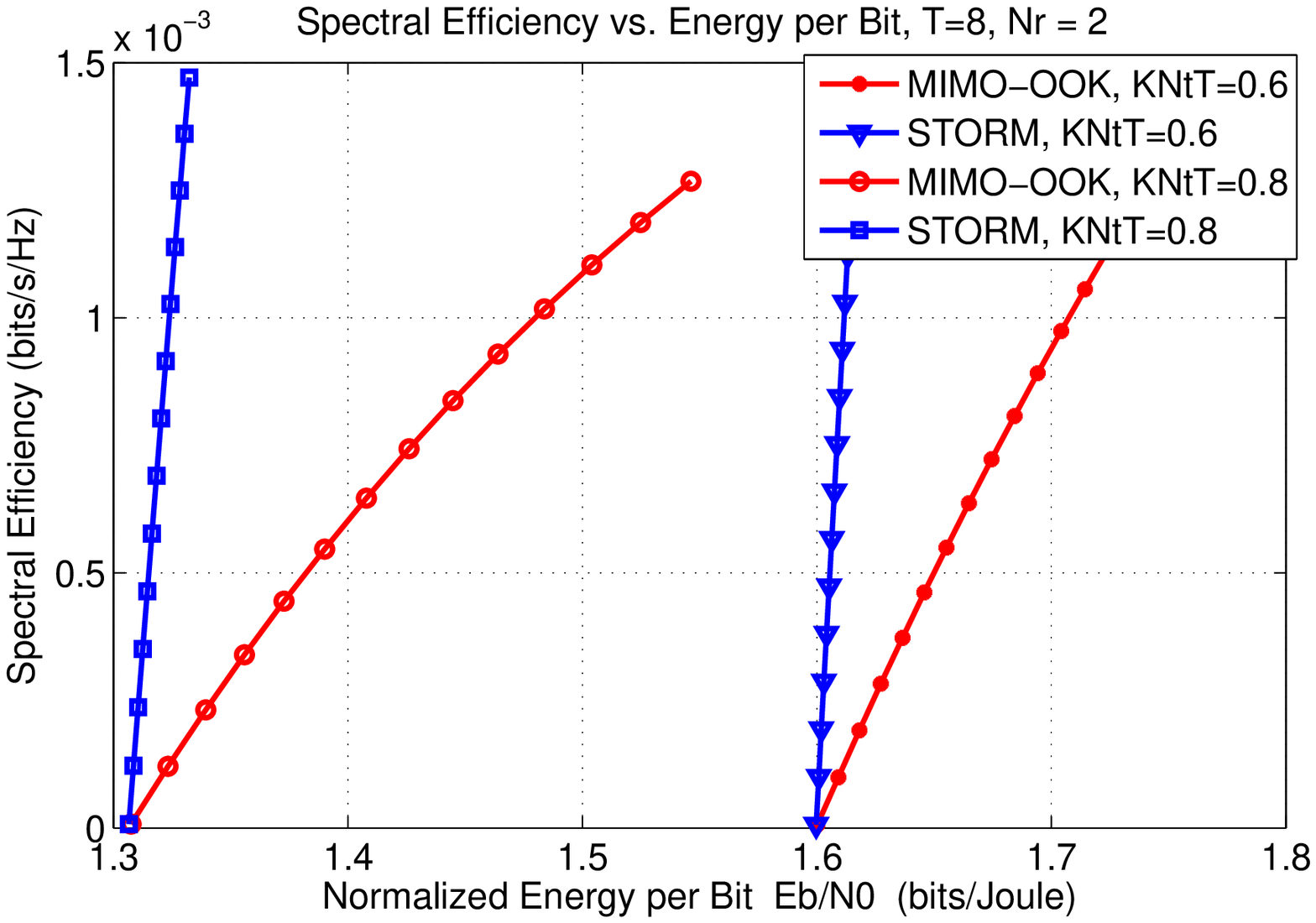}}}
    \caption{Plot of spectral efficiency vs. energy/bit of STORM using the second order approximation of $I(P)$ in (\ref{STORMTaylor}),  for different values of $KN_tT$. }  \label{STORM_SpecEff_KNtT} 
  \end{figure}

\begin{figure}
\centerline{\scalebox{0.65}{\includegraphics{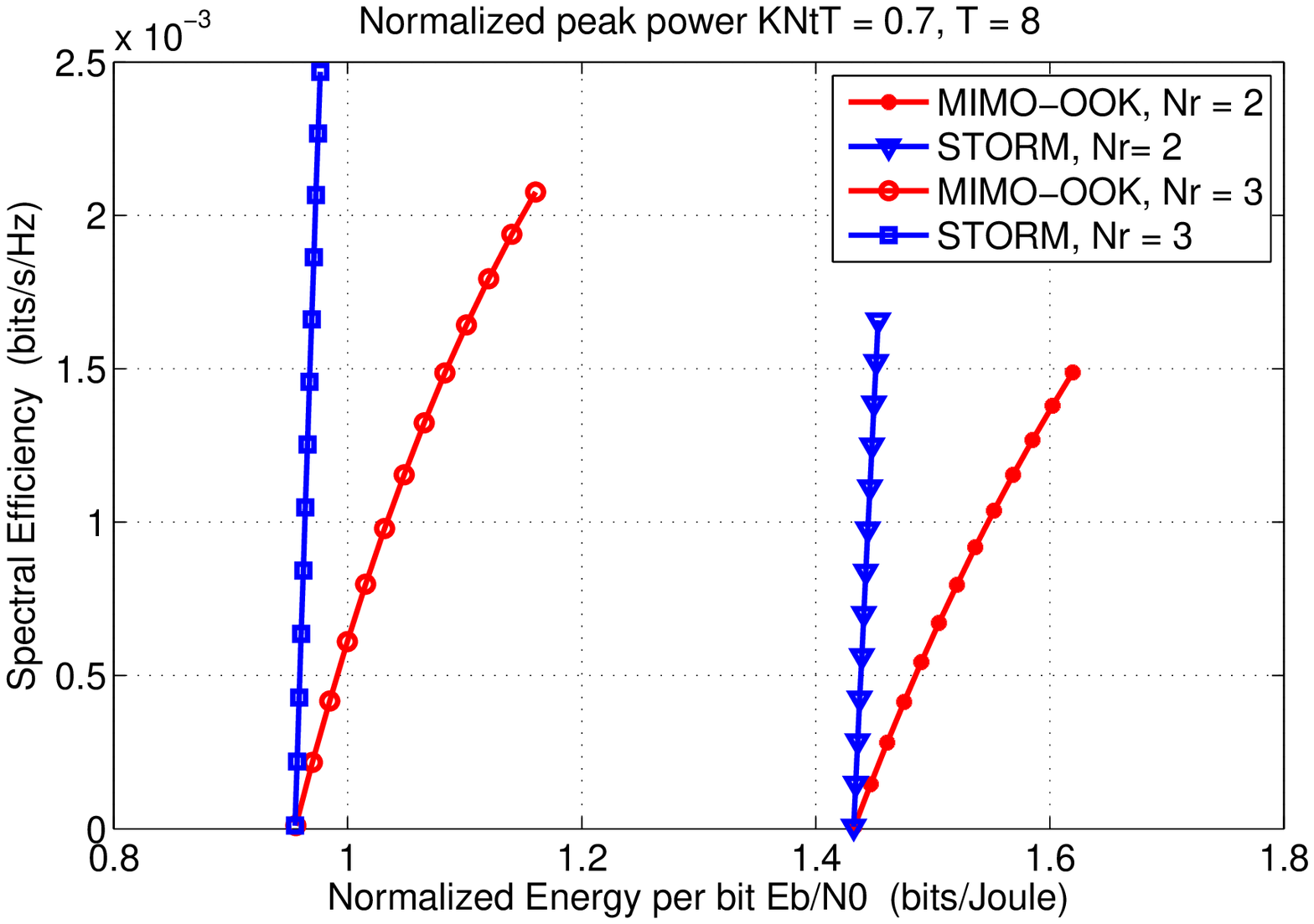}}}
    \caption{Plot of spectral efficiency vs. energy/bit of STORM using the second order approximation of $I(P)$ in (\ref{STORMTaylor}),  for different values of $N_r$. }  \label{STORM_SpecEff_Nr} 
\end{figure}

\begin{figure}
  \centerline{\scalebox{0.65}{\includegraphics{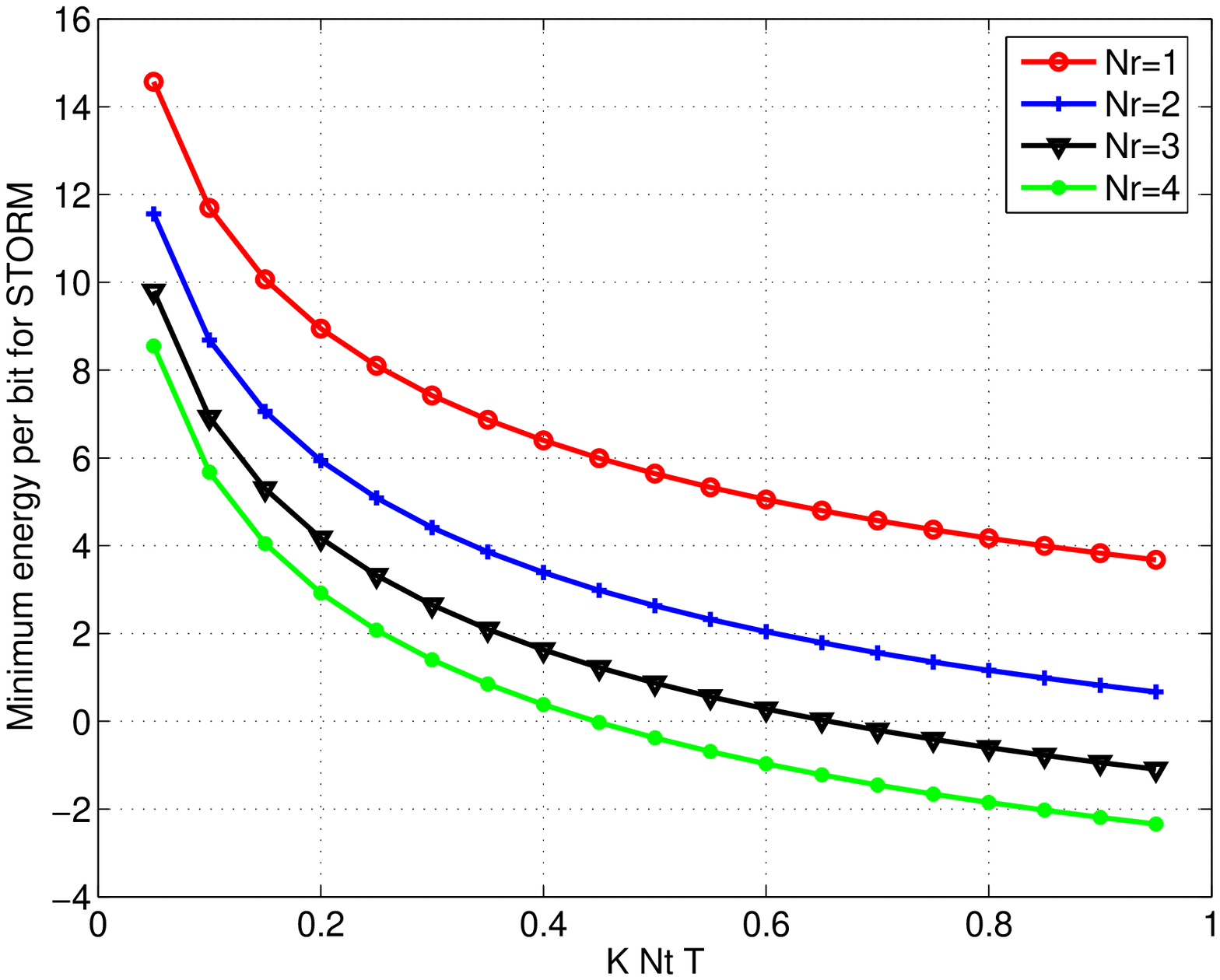}}}
    \caption{Energy per bit (dB) of STORM vs.  $K N_t T$  for different $N_r$} \label{STORMEbmin} 
  \end{figure}

\begin{figure}
  \centerline{\scalebox{0.65}{\includegraphics{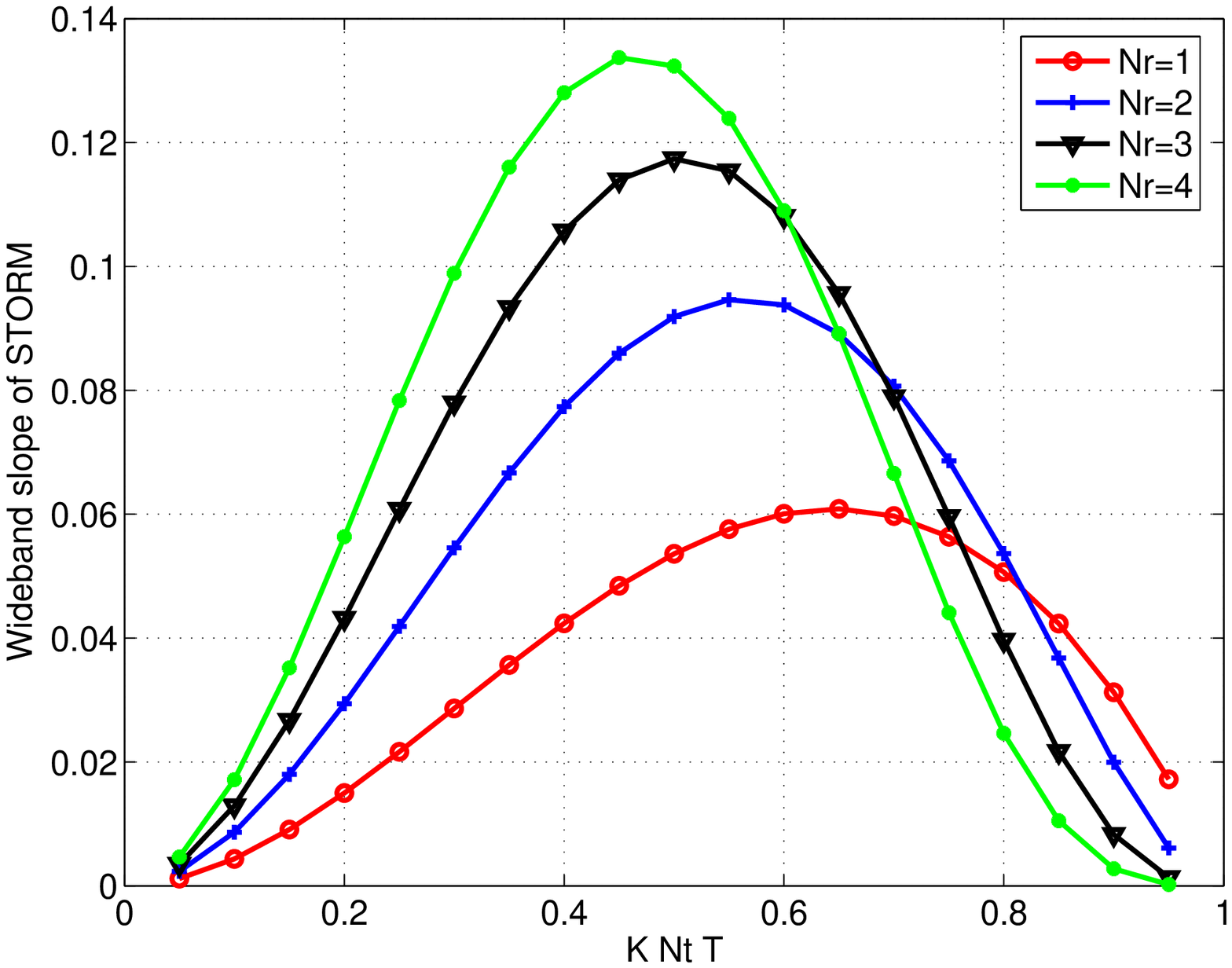}}}
    \caption{Wideband slope of STORM vs. $K N_t T$ for different $N_r$} \label{STORMWBslope} 
  \end{figure}

\begin{figure}
  \centerline{\scalebox{0.65}{\includegraphics{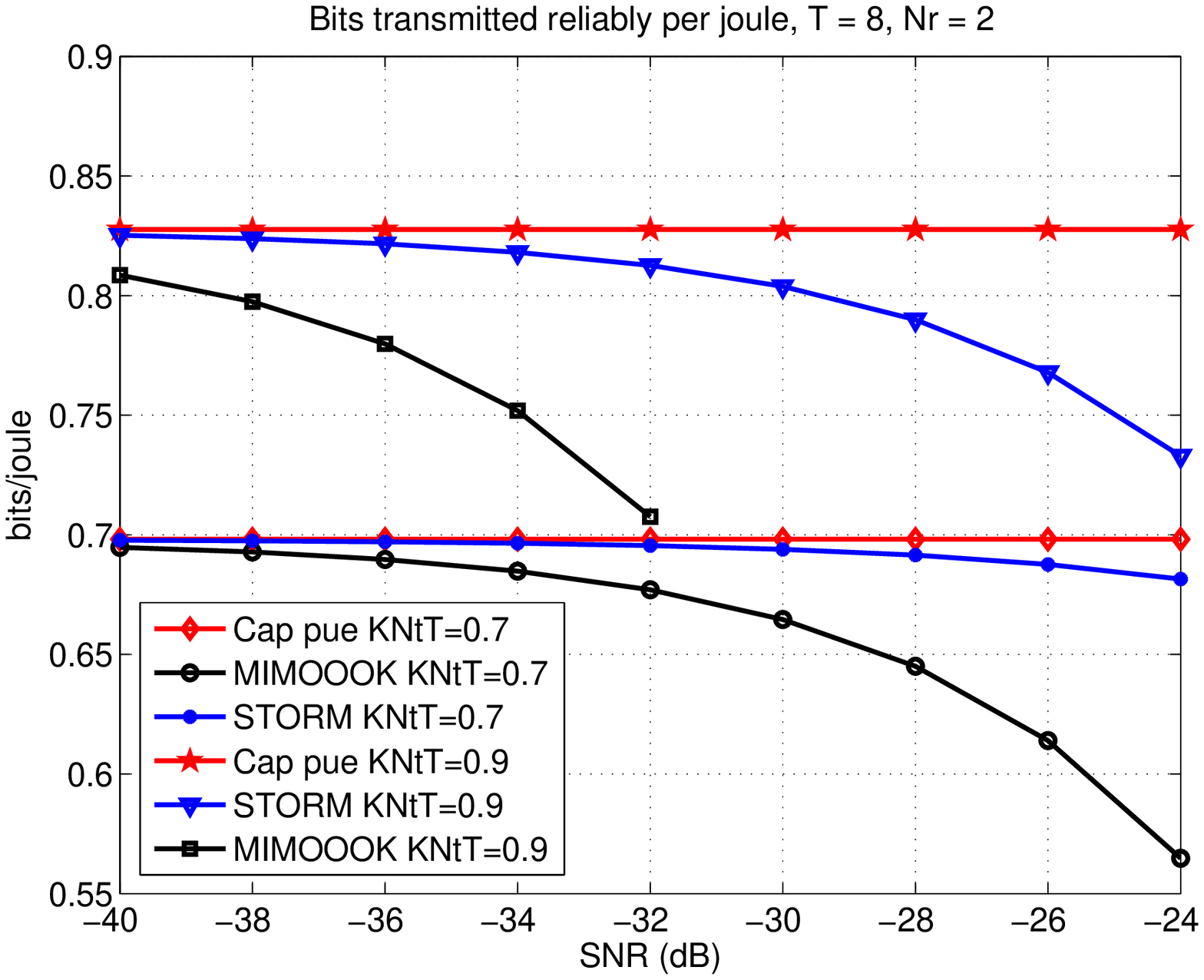}}}
    \caption{Plot comparing first order approximation of $\frac{I(P)}{P}$ in (\ref{STORMTaylor}) vs. $P$ for STORM and MIMO-OOK for different values of $K N_t T$. } \label{STORMvsMIMOOOK} 
  \end{figure}

\begin{enumerate}
\item{} From (\ref{STORMcapune}), it is seen that $ \lim_{K \to \infty} \dot{C}(0) = N_r $. Therefore, for asymptotically large peak-powers, the well known result on the capacity per unit energy with only an average power constraint \cite{VerduS:efficiency}  which is common to both coherent and noncoherent MIMO channels, is recovered. Indeed, when $N_r = 1$, we obtain the minimum energy to transmit one bit of information to be $-1.59$ dB, which is a classical result. By relaxing the peak-constraint, STORM can be seen to be optimal even for the case when there is merely an average power constraint (or with respect to infinite bandwidth capacity). 

\item{} When the signals are just subject to an average power constraint, it is shown in \cite{VerduS:efficiency}  that ${\mathcal S}_0=0$ for the noncoherent MIMO channel. Therefore,  signals whose energy per bit approaches ${\frac{E_b}{N_0}}_{min}$ would have to have bandwidths that become prohibitively large. However, when there is an additional peak-power constraint $K$ which is a fixed constant, and for the case when the normalized peak power $KN_tT < 1$, Corollary \ref{STORMSTORMEbmin} shows that ${\mathcal S}_0$ is strictly positive. Hence, it is realistic to design signals that achieve the ${\frac{E_b}{N_0}}_{min}$ in this scenario for low but non-vanishing SNR. Similar insights were also noted in \cite{GursoyMC:nonc} but in the simpler context of the SISO Rician fading channel with unit block length under peak and average power constraints. 

\item{} While both MIMO-OOK and STORM achieve ${\frac{E_b}{N_0}}_{min}$, according to Corollary \ref{STORMSTORMEbmin}, the wideband slope of STORM is higher by a factor of $T$.  This means that at a certain energy per bit and for the same transmission rate, and as ${\rm SNR} \to 0$, the bandwidth needed by STORM for the same spectral efficiency is less than that of MIMO-OOK by a factor of $T$. Given typical values of the coherence time $T$, this higher  spectral efficiency of STORM can translate into huge savings. To give a sense of the significant gains, Figures \ref{STORM_SpecEff_KNtT} and \ref{STORM_SpecEff_Nr} plot the spectral efficiency vs. the energy per bit for STORM and MIMO-OOK.

\item{} Figures \ref{STORMEbmin} and \ref{STORMWBslope} plot the energy per bit and wideband slope of STORM vs. the normalized peak power $KN_tT$, for different values of $N_r$. As the normalized peak power  increases, it is seen that the $\frac{E_b}{N_0}_{min}$ decreases. This is expected as peakier signaling is more energy efficient. However, as the normalized peak power gets close to $1$, the wideband slope approaches $0$. In fact, the wideband slope attains its maximum at an intermediate value between $0$ and $1$ (say $K N_t T = c^{*}$). Since for any point in the region $0 \leq K N_t T \leq c^{*}$ there is a point corresponding to $c^{*} \leq K N_t T \leq 1$ with lower $\frac{E_b}{N_0}_{min}$ and the same wideband slope, it makes most sense to operate in the region $c^{*} \leq K N_t T \leq 1$.  Assuming only an average power constraint, the analysis in  \cite{VerduS:efficiency} shows that ${\mathcal S}_0 = 0$ for noncoherent communications. The scheme that achieves the $\frac{E_b}{N_0}_{min}$ has the non-zero signals migrating to $\infty$  in amplitude as $P \to 0$. The results in \cite{VerduS:efficiency} show in effect that it is unrealistic to realize the peak-unconstrained minimum energy per bit (STORM having zero wideband slope for all $K N_t T \geq 1$ is clearly a stronger statement). Under realistic assumptions on the peak-constraint however, it has been shown here that ${\mathcal S}_0 > 0$ is possible when $K N_t T < 1$. Moreover, a sharp characterization is provided which shows that there is a tradeoff between  $\frac{E_b}{N_0}_{min}$ and ${\mathcal S}_0$ for STORM in the region $c^{*} \leq K N_t T \leq 1$.

\item{} For the same number of bits transmitted reliably per joule at low SNR, MIMO-OOK requires an operating SNR which is $10 \log_{10} T$ dB smaller than that of STORM. This can be seen from the fact that the wideband slope of STORM is $T$ times that of MIMO-OOK and that mutual information per joule is given as
\begin{equation}
 \frac{I(P)}{P} = \dot{I}(0)(\log_2 e)   + \frac{1}{2}\ddot{I}(0)(\log_2 e) P  + o(P). \label{STORMTaylor} 
 \end{equation}
 and the wideband slope is $ {\mathcal S}_0^{I} = \frac{2\left[\dot{I}(0)\right]^2}{-\ddot{I}(0)} $. Now, since the peak-power is a fixed constant, this implies that the PAPR of  MIMO-OOK at any small but non-vanishing SNR would be greater than that of STORM by a factor of $T$.  Since in the low SNR regime, peakiness of the signal constellations is a crucial factor, using STORM can  potentially result in large reductions in the required PAPR and facilitate implementation. These large savings are illustrated in Figure \ref{STORMvsMIMOOOK}, where the approximation of $\frac{I(P)}{P}$ vs. $P$ is plotted for STORM and MIMO-OOK. In the example shown, the convergence to the capacity per unit energy is faster for STORM by a factor of $10\log_{10} T = 9$ dB relative to MIMO-OOK. 


\item{}  It has been shown in Corollary \ref{STORMCorollaryWBSlope} that whenever $KN_tT > 1$, the wideband slope is $0$. Therefore, even though the noncoherent capacity per unit energy is $N_r \log_{2} e$ bits/joule, it is prohibitively expensive (in terms of bandwidth) to reliably transmit at any rate more than the peak-constrained capacity per unit energy evaluated at $KN_tT = 1$ which from equation (\ref{STORMcapune}) is $N_r (\log_2 e  - 1)$ bits/joule. Hence, the capacity per unit energy at $KN_tT = 1$ can be taken to be the realistic limit for noncoherent MIMO communication. Note that this limit is also $N_r$ bits/joule smaller than the coherent capacity per unit energy. Since the analysis of the noncoherent channel neither assumes any particular scheme for channel estimation nor does it ignore the resources for (implicit) channel estimation, the realistic capacity per unit energy of $N_r (\log_2 e  - 1)$ bits/joule can be argued as being more fundamental than the coherent capacity per unit energy of $N_r \log_{2} e$ bits/joule. The difference between the two can be thought of as the fundamental or minimal cost of (implicit) channel estimation. 

\item{} The dependence of $\frac{E_b}{N_0}_{min}$ on $K$, $N_t$ and $T$ is only through the product $KN_tT$. So, increasing one or more these quantities has the effect of lowering $\frac{E_b}{N_0}_{min}$. However, this effect is beneficial when $KN_tT <c^*$ and beyond that the tradeoff between energy efficiency and bandwidth efficiency is quantified here that allows a designer to choose a suitable operating point. To illustrate this point, Figure \ref{convergence_withP} plots the approximation of $\frac{I(P)}{P}$ vs. $P$ for different values of $KN_tT$. It is evident from Figure \ref{convergence_withP}, that even as $KN_tT$ gets close to one, the bits required to transmit reliably converges to the capacity per unit energy at much smaller SNRs (and hence larger bandwidths). Since the PAPR of STORM at SNR $P$ is $\frac{KN_t}{P}$,  it interesting to note that when $KN_tT$ is fixed, increasing $T$ decreases the $PAPR$ required for the same energy per bit which is an advantage in practice. Increasing $N_t$ with $KN_tT$ fixed, decreases $K$ and therefore reduces the peak-power per antenna and time slot (though not changing the PAPR), which may also be helpful in practice.  

\item{} An interesting observation from Figures \ref{STORMEbmin} and \ref{STORMWBslope} is that using more receive antennas always lowers $\frac{E_b}{N_0}_{min}$, while it does not always increase the wideband slope. Figure \ref{STORM_Nr} illustrates that while the approximation of $\frac{I(P)}{P}$ increases with $N_r$ in general, the convergence to the capacity per unit energy occurs more slowly and hence a lower SNR is needed to operate close to it as $N_r$ increases.  

\item{} Even though the optimal scheme for a cardinality more than $T+1$ is yet unknown, STORM offers a concrete solution whose structure is also simple and practical. In \cite{VerduS:efficiency}, the positive impact on the wideband slope of using constellations with cardinality greater than two is illustrated via several contexts other than under the noncoherent assumption. Even so, following  \cite{VerduS:unitcost} and due to analytical convenience, many recent papers \cite{GursoyMC:nonc, SethuramanV:cuelowsnr, WuX:MIMOexp} in noncoherent communications focus on the two point ON-OFF scheme to achieve the capacity per unit energy. The results in this section  demonstrate that there are  compelling reasons to look beyond the two point ON-OFF scheme in the low SNR regime.

\item{} Recently, \cite{ZhengL:chcoh, Rays:noncWB, RaghavanV:sparse, HariharanG:ITA2007} have investigated the possibility of channel coherence length scaling with SNR, so as to diminish the cost of acquiring channel knowledge. It should be interesting to pose and solve the optimization problems of this work under such scenarios.

  \end{enumerate}

 \section{Conclusion}

We pose two important problems on reliable communications over noncoherent MIMO spatially i.i.d. Rayleigh fading channels at low SNR. In both formulations, we assume an average-power constraint on the input and a natural per-antenna, per-time slot  peak-power constraint. In the first problem formulation, the peak-power to average-power ratio is held fixed (PPAPR-constrained) and the mutual information which grows as $O({\rm SNR}^2)$ is maximized up to second order jointly over input signal matrices and their respective probabilities, when the cardinality of the constellation is no greater than $T+1$ ($T$ is the coherence blocklength). In the second problem formulation (peak-constrained), the peak-power is a fixed constant independent of SNR. Here, necessary and sufficient conditions for a constellation of any cardinality to achieve the minimum energy/bit are derived. Over the set of all $T+1$ point constellations which achieve the minimum energy/bit, we optimize the second order behaviour of mutual information. The resulting constellations are  both  first and second order optimal among all $T+1$ point constellations. Both the PPAPR-constrained and peak-constrained problems result in finite dimensional non-convex optimization problems. Even so, they admit elegant solutions in closed form, which are identical in both formulations. We refer to this common solution as Space Time Orthogonal Rank-one Modulation (STORM), and it provides several new insights on noncoherent   communications at low SNR. 

In the PPAPR-constrained case, we show that the $T+1$ point STORM is near-optimal with respect to the maximum mutual information up to second order with unconstrained cardinality even for modest values of $T$ and PAPR. Therefore, there is not much to be gained by using more than $T+1$ points in the PPAPR-constrained case.   In the peak-constrained case, our approach enables us to provide a sharp characterization of the first and second order behavior of noncoherent MIMO capacity, that also sheds light on the cost of implicit estimation of channel state information in the low SNR regime. The energy/bit and the wideband slope achieved by STORM also reveals a fundamental energy-vs-bandwidth efficiency tradeoff that enables the determination of the operating (low) SNR and peak power most suitable for a given application. Moreover, while the more conventional MIMO On-Off Keying (OOK) also achieves the minimum energy per bit, STORM has a wideband slope that is $T$ times greater which translates into an increase in bandwidth efficiency (or a decrease in the PAPR) by a factor of $T$ in the wideband regime. Given typical values of the coherence blocklength $T$, these gains are potentially huge.

\section{Acknowledgements} \nonumber
The authors would like to thank the anonymous reviewers for their helpful comments on the originally submitted version of this paper.
\begin{appendix}

\subsection{Proof of non-convexity} \label{STORMnonconvexity}
A simple argument is given to show that (\ref{STORMnegzero}) is a non-convex optimization problem.

   We need the following definition of matrix convexity.
   \begin{definition} 
   A function $f : \Re^{n \times n} \rightarrow \Re^{m \times m}$ is matrix convex  with respect to  matrix inequality if for any positive semidefinite $\Xb_1 , \Xb_2$ and for any $\theta \in [0,1]$
   \begin{eqnarray}
   f\left(\theta \Xb_1 + (1-\theta) \Xb_2\right) &\preceq& \theta f\left(\Xb_1\right) + (1 - \theta) f\left(\Xb_2\right) \ .
   \end{eqnarray}
   \end{definition}


Since $\{\Xb_i\}_{i=1}^{L}$ is a set of complex matrices, the optimization over the signals amounts to an equivalent joint optimization over the real and imaginary parts of $\Xb_i$ given by $\Xb_i = \widehat{\Xb}_i + j \widetilde{\Xb}_i \ , \ \ \forall i$. In order to show that this joint optimization is non-convex, we will consider the contour given by $\widetilde{\Xb}_i = \Zrb \ , \ \ \forall i$. With the imaginary parts being zero, the function in (\ref{STORMnegzero}) becomes  $\sum_i P_i\left(1-P_i\right)\tr \left(\widehat{\Xb}_i\widehat{\Xb}_i^{*}\widehat{\Xb}_i\widehat{\Xb}_i^{*}\right)$

 It can be seen that $g(\widehat{\Xb})=\widehat{\Xb}\widehat{\Xb}^{*}$ is matrix-convex over $\widehat{\Xb}$, and $h(\Ab)=\tr(\Ab\Ab^{*})$ is a non-decreasing convex function over positive semidefinite matrices $\Ab$. Therefore, the composition $f(\widehat{\Xb}) = h \circ g = \tr \left(\widehat{\Xb}\widehat{\Xb}^{*}\widehat{\Xb}\widehat{\Xb}^{*}\right)$ is a convex function over $\widehat{\Xb}$ \cite{BoydS:Optim}. Further, since $\tr(\widehat{\Xb}\widehat{\Xb}^{*})$ and $\|\widehat{\Xb}\|_{\infty}$ are convex functions of $\widehat{\Xb}$ \cite{BoydS:Optim},  the constraints $\sum_{i} P_i \tr(\widehat{\Xb}_i\widehat{\Xb}_i^{*}) \leq E$  and $\|\widehat{\Xb}_i\|_{\infty} \leq \sqrt{K}$  are  convex sets in $\{\widehat{\Xb}_i\}_{i=1}^{L}$.  For an arbitrary but fixed set of probabilities $\{P_i\}_{i=1}^{L}$, the objective function is convex in $\{\widehat{\Xb}_i\}_{i=1}^{L}$, while the constraint set is the intersection of convex sets and is hence convex. Therefore, the problem of optimizing (\ref{STORMnegzero}) over $\{\widehat{\Xb}_i\}_{i=1}^{L}$  is a convex maximization problem and not a convex optimization problem. Since for a fixed $\{P_i\}_{i=1}^{L}$, the problem of optimizing over  $\{\Xb_i\}_{i=1}^{L}$ is a non-convex optimization problem for the imaginary parts of $\Xb_i$ fixed, the joint optimization over $\{P_i\}_{i=1}^{L}$ and $\{\Xb_i\}_{i=1}^{L}$ is also non-convex.

 \subsection{A low complexity block decoder} \label{STORMFFT}

 In some applications, decoding of a block of symbols at a time may be required. This need arises for instance in uncoded systems, where there is no coding across blocks. Another possibility is when there is coding across blocks, but hard decision decoding is employed at the receiver so that the blocks of symbols are first decoded via the MAP rule following which the outer code is decoded. In all such cases, we show in this section that the optimal MAP decoding of STORM  can be simplified  using Fast Fourier Transform (FFT) or Fast Hadamard Transform (FHT) algorithms.

  Consider the $T+1$ point STORM as described in (\ref{STORMnonzeropoints}) and (\ref{STORMlargepeakopt}). Let the received signal matrix be $\Rb \in \Cplx^{T \times N_r}$. The optimal MAP rule to decode a block at the receiver is 
 \begin{eqnarray}
 \widehat{j}&=& \max_j \ \ \  P_j \ p\left(\Rb | \Xb_j\right) \\
 &=& \max_j \ \ \  P_j \  \frac{\exp\left\{-\tr \left(\Rb^{*}\left(\Ib_{T} +  \Xb_j\Xb_j^{*}\right)^{-1}\Rb\right)\right\}}{\pi^{TN_r}\left|\Ib_{T} +  \  \Xb_j\Xb_j^{*}\right|^{N_r}} \label{STORMdetection}
 \end{eqnarray}
 For convenience, we will first find the maximum in (\ref{STORMdetection}) among the non-zero signal matrices, and then compare it with the metric for the zero matrix. Substituting STORM that is defined with permutation matrix $\Pb$, we get that the maximum metric among non-zero matrices is 
 \begin{eqnarray}
 \max_{i = 1, \ \dots \ , T} \ \ \   \frac{E}{ K N_t T^2} \  \frac{\exp\left\{-\tr \left(\Yb^{*}\left(\Ib_{T} +  KN_t \ \vb_i  \vb_i^{*} \right)^{-1}\Yb\right)\right\}}{\pi^{TN_r}\left|\Ib_{T} +    KN_t \ \vb_i \vb_i^{*}\right|^{N_r}} \ ,\label{STORMinterdet}
 \end{eqnarray} 
 where $\Yb = \Pb^{*}\Rb$ is a sufficient statistic, which is simply the received matrix with the permutation removed.  The term $\left(\Ib_{T} +  KN_t \ \vb_i  \vb_i^{*} \right)^{-1}$ can be simplified by applying the Woodbury's identity, i.e., using 
\begin{eqnarray*}
\left(\Ab + \Bb\Cb\Db\right)^{-1} = \Ab^{-1} - \Ab^{-1}\Bb\left(\Cb^{-1} + \Db\Ab^{-1}\Bb\right)^{-1}\Db\Ab^{-1} \ .
\end{eqnarray*}
Also, using the identity $\left|\Ib + \Ab \Bb\right| = \left|\Ib + \Bb \Ab\right|$,  (\ref{STORMinterdet}) becomes
 \begin{eqnarray}
 \max_{i = 1, \ \dots \ , T} \ \ \  \frac{E \ \ \exp \left\{-\tr(\Yb\Yb^{*})\right\}}{\pi^{T N_r} K N_t T^2 (1 + K N_t T)^{N_r}} \ \exp \left\{\frac{K N_t}{1+K N_tT} \tr \left(\Yb^{*} \vb_i\vb_i^{*}\Yb \right)\right\} \ . \label{STORMnonzerometric}
 \end{eqnarray}
 Clearly, among the non-zero constellation matrices, the MAP metric is maximized when $\|\Yb^{*} \vb_i\|^2$ is maximized. Let $\Vb$ be the $T$ dimensional DFT or Hadamard matrix. Then each row of the matrix $\Yb^{*} \Vb$ would represent the DFT or Hadamard transform of the corresponding row of $\Yb^{*}$. The non-zero constellation matrix with the maximum MAP metric would therefore correspond to the column of $\Yb^{*} \Vb$ with the maximum $l_2$-norm. The $N_r$ DFTs or Hadamard transforms involved can be efficiently computed using fast algorithms (FFTs and FHTs). Now, the metric corresponding to the zero matrix would be
\begin{eqnarray}
\left(1 - \frac{E}{KN_tT}\right) \frac{\exp{\left(-\tr(\Yb\Yb^{*})\right)}}{\pi^{TN_r}} \ . \label{STORMzerometric}
\end{eqnarray} 
Since this is a constant for a given received signal, we can divide the metric in (\ref{STORMnonzerometric}) by (\ref{STORMzerometric}) and then take the natural logarithm of the resulting expression so that
\begin{equation}
\Omega_i =\ln \left\{\frac{E}{T (KN_tT - E) (1 + KN_tT)^{N_r}}\right\} + \frac{KN_t}{1 + KN_tT} \tr\left(\Yb^{*}\vb_i\vb_i^*\Yb\right) \ .
\end{equation}
Now letting $i = \arg \max_{k = 1, \dots , T} \ \|\Yb^{*} \vb_k\|^2 $, the final simplified decoding rule can be given as
\begin{equation} 
\widehat{j}= \left\{ \begin{array}{ll}
 i & {\rm if} \; \Omega_i \geq 0 \\
            T+1 & {\rm if} \; \Omega_i < 0 .
            \end{array} \right.
\end{equation}

\subsection{Derivation of MIMO-OOK} \label{STORMmimoook}

 \begin{theorem} \label{STORMCUEONOFF}
 The capacity per unit energy (in nats/joule) for the i.i.d. MIMO block Rayleigh fading channel with a peak power constraint on the input signal $\|\Xb\|_{\infty} \leq \sqrt{K}$ is
 \begin{eqnarray}
  \dot{C}(0) = N_r \left(1- \frac{ \log\left(1 + KN_tT\right)}{KN_tT} \right)  \ ,
 \end{eqnarray}
 and is achieved as $P \to 0$ by the two point constellation given as
 \begin{eqnarray}
 \left(\Xb_1 , P_1\right)&=& \left(\sqrt{K} \ \vb \  \wb^{*}  \ , \ \frac{P}{KN_t}\right) \label{STORMMIMOOOKnonzero}\\
 \left(\Xb_2 , P_2\right)&=& \left(\Zrb_{T \times N_t} \ , \ 1 - \frac{P}{KN_t}\right) \ , \label{STORMMIMOOOKzero}
 \end{eqnarray} 
 where $\vb \in \Cplx^{T \times 1}$, $\wb \in \Cplx^{N_t \times 1}$ and $|\left[\vb \wb^{*}\right]_{mn}| = 1 \ \forall \ i, m, n$.
 \end{theorem}
  \proof
 From \cite{VerduS:unitcost}, it is known that to achieve the channel capacity per unit energy, it is enough to transmit one non-zero symbol, given in (\ref{STORMMIMOOOKnonzero}), apart from the  symbol $\Zrb$. Since we are dealing with a memoryless, discrete and matrix input channel (\ref{STORMsyseqn1}) with the cost-function given by $b(\Xb) = \tr(\Xb\Xb^{*})$, the capacity per unit energy under a fixed peak power constraint is given by \cite{VerduS:unitcost}
  \begin{eqnarray}
   \dot{C}(0)&=& \sup_{\Xb \neq \Zrb \atop {\|\Xb\|_{\infty} \leq \sqrt{K}}}  \frac{D\left(p\left(\Yb | \Xb\right) \ || \ p\left(\Yb | \Zrb\right)\right)}{\tr(\Xb\Xb^{*})} \ .
  \end{eqnarray}
  Using the expression for the Kullback-Liebler distance which can be obtained easily (c.f. \cite{BorranMJ:NoncSTC}), we obtain
  \begin{eqnarray}
   \dot{C}(0)&=&\sup_{\Xb \neq \Zrb \atop {\|\Xb\|_{\infty} \leq \sqrt{K}}} \ N_r \left(1- \frac{ \log \det \left(\Ib_T + \Xb\Xb^{*}\right)}{\tr(\Xb\Xb^{*})} \right) \\
  &=&\sup_{\Xb \neq \Zrb  \ \ , \ d \atop { \tr(\Xb\Xb^{*}) = d \ , \ \|\Xb\|_{\infty} \leq \sqrt{K}} }  \ N_r \left(1- \frac{ \log \det \left(\Ib_T + \Xb\Xb^{*}\right)}{d} \right). \label{STORMUBCue1}
  \end{eqnarray}
  Let the matrix $\Xb\Xb^{*}$ have eigenvalues $\left\{\lambda_i\right\}_{i=1}^{T}$. Then (\ref{STORMUBCue1}) can be upper bounded as
  \begin{eqnarray}
   \dot{C}(0) &\leq& \sup_{\{\lambda_i\}_{i=1}^{T} \ , \ d \atop {\sum_i \lambda_i  = d \ , \ d \leq KN_tT} } \  N_r \left(1- \frac{ \sum_i \log(1 + \lambda_i) }{d} \right) \label{STORMUBCue2} \\
  &=&\sup_{ \ d \atop { d \leq KN_tT} } \  N_r \left(1- \frac{ \log(1 + d) }{d} \right) \label{STORMUBCue3}\\
  &=& N_r \left(1- \frac{ \log\left(1 + KN_tT\right)}{KN_tT} \right) \ . \label{STORMUBCue4}
  \end{eqnarray}
  The expression in (\ref{STORMUBCue3}) is obtained by noting that since  $-\sum_i \log(1 + \lambda_i)$ is a convex function of \newline $\left[\lambda_1 \ \lambda_2 \ \ \dots \ \lambda_T\right]^{T}$, the supremum in (\ref{STORMUBCue2}) is achieved at the extreme point $\left[d \ \ 0 \ \ \dots \ \ 0\right]^T$ by Lemma \ref{STORMextreme}. Since $\left(1- \frac{ \log(1 + d) }{d} \right)$ is a monotonically increasing function of $d$, we obtain (\ref{STORMUBCue4}) by substituting the maximum value of $d$. The inequality in (\ref{STORMUBCue2}) is achieved with equality when $\Xb$ is of unit rank, $\tr(\Xb\Xb^{*})=d$ and $\|\Xb\|_{\infty} \leq \sqrt{K}$. The supremum in (\ref{STORMUBCue3}) is achieved when $d=KN_tT$, and  the unit rank $\Xb$ satisfies both $\tr(\Xb\Xb^{*})=KN_tT$ as well as $\|\Xb\|_{\infty} \leq \sqrt{K}$ which in turn is true iff it is of the form given in (\ref{STORMMIMOOOKnonzero}).
To satisfy the average power constraint, set $P_1=\frac{P}{KN_t}$.   \endproof


\subsection{Proof of Theorem \ref{STORMproppearson}}  \label{STORMproofproppearson}

The results regarding generalized on-off signaling given in \cite{VerduS:efficiency} are employed. In particular, note that Theorem 10 in \cite{VerduS:efficiency} provides the ${\frac{E_b}{N_0}}_{min}$ and ${\mathcal S}_0$ achieved by a generalized on-off signaling scheme. For convenience, that result is summarized here. 

The generalized on-off signaling scheme has  a $P_0$ mass at the all-zero matrix $\Zrb_{T \times N_t}$. The input pdf conditioned on the input being nonzero is denoted by $\overline{P}_{X}$, with distribution $\overline{F}_{X}$. With the input pdf conditioned on the all-zero matrix given by ${P}^{\Zrb}$, the input pdf is
\begin{eqnarray}
P_{X}=(P_0){P}^{\Zrb} + (1-P_0) \overline{P}_{X} \ .
\end{eqnarray}
Denoting the pdf of the output conditioned on the input by $P_{Y | X}$, the output pdf corresponding to $\overline{P}_{X}$ is given by 
\begin{eqnarray}
\overline{P}_{Y} = \int P_{Y | X=\Xb} \ d\overline{F}_{X}(\Xb)\ . 
\end{eqnarray}
The wideband slope  ${\mathcal S}_0$ achieved by generalized on-off signaling is
\begin{eqnarray}
 {\mathcal S}_0&=&\frac{2}{T}\frac{\left(\E_{\overline{P}_{X}}\left[D\left(P_{Y | X} || P_{Y | X =\Zrb}\right)\right]\right)^2}{\Delta(\overline{P}_{Y}||P_{Y | X =\Zrb})} \ , \label{STORMs0}
 \end{eqnarray}
where $\Delta(.||.)$ denotes the Pearson's $\chi$-divergence and is defined as
\begin{eqnarray}
\Delta(\overline{P}_{Y}||P_{Y | X =\Zrb}) \define \E_{P_{Y | X =\Zrb}}\left[\left(\frac{\overline{P}_{Y}}{P_{Y | X =\Zrb}} - 1\right)^2\right] \ . \label{STORMPearsons}
\end{eqnarray}
For the channel model under consideration in this paper, we have
\begin{eqnarray*}
&\overline{P}_{Y}& =\sum_{i=1}^{M} \frac{P_i}{(1-P_0)} \frac{1}{\pi^{TN_r} \left|\Ib + \Xb_i\Xb_i^{*}\right|^{N_r}} \exp \left\{-\tr\left(\Yb^{*}\left(\Ib+\Xb_i\Xb_i^{*}\right)^{-1} \Yb\right)\right\} \\
& P_{Y | X=\Zrb}&= \frac{\exp(-\tr(\Yb^{*}\Yb))}{\pi^{TN_r}},
\end{eqnarray*}
and using the above expressions in (\ref{STORMPearsons}), one obtains
\begin{eqnarray}
\Delta\left(\overline{P}_{Y}||P_{Y | X =\Zrb}\right)&=&\E_{P_{Y | X=\Zrb}}\left[ \sum_{i=1}^{M} \frac{P_i^2}{(1-P_0)^2} \frac{e^{2 \tr\left(\Yb^{*}(\Ib - (\Ib + \Xb_i\Xb_i^{*})^{-1})\Yb\right)}}{\left|\Ib+\Xb_i\Xb_i^{*}\right|^{2N_r}}  \right.   \nonumber \\
&& \hspace{15mm} + 2 \sum_{i, j\neq i} \frac{P_i P_j}{(1-P_0)^2} \frac{e^{\tr\left(\Yb^{*}\left(2\Ib - (\Ib+\Xb_i\Xb_i^{*})^{-1}  - (\Ib+\Xb_j\Xb_j^{*})^{-1}\right)\Yb\right)}}{\left|\Ib+\Xb_i\Xb_i^{*}\right|^{N_r}\left|\Ib+\Xb_j\Xb_j^{*}\right|^{N_r}} \nonumber \\
&& \hspace{15mm} \left.  - 2 \sum_{i=1}^{M} \frac{P_i}{1-P_0} \frac{e^{\tr\left(\Yb^{*}\left(\Ib - (\Ib+ \Xb_i\Xb_i^{*})^{-1}\right)\Yb\right)}}{\left|\Ib+\Xb_i\Xb_i^{*}\right|^{N_r}} + 1 \right]
\label{STORMDeltainter1}
\end{eqnarray}
The above expression can be evaluated using the result from \cite{TurinGL:quad} that if $\zb$ is $\NCplx(\Zrb,\Kb)$ distributed, then $\E_{\zb}\left[\exp\left(\zb^{*}\Ab\zb\right)\right] =\left\{\det\left(\Ib - \Kb \Ab \right)\right\}^{-1}$ if $\Ib - \Kb \Ab$ is positive definite. Otherwise, the expectation diverges. Hence (\ref{STORMDeltainter1}) becomes
\begin{eqnarray}
\Delta\left(\overline{P}_{Y}||P_{Y | X =\Zrb}\right)&=& 1 \ \ + \ \ \sum_{i=1}^{M} \frac{P_i^2}{(1-P_0)^2} \frac{1}{\left|\Ib+\Xb_i\Xb_i^{*}\right|^{2N_r}\left|\Ib - (2\Ib - 2(\Ib + \Xb_i\Xb_i)^{-1})\right|} \nonumber \\
&& \hspace{-12mm}  + \sum_{i \atop j \neq i} \frac{2  P_i P_j}{(1-P_0)^2} \frac{1}{\left|\Ib+\Xb_i\Xb_i^{*}\right|^{N_r}\left|\Ib+\Xb_j\Xb_j^{*}\right|^{N_r} \left|(\Ib+\Xb_i\Xb_i^{*})^{-1} + (\Ib+\Xb_j\Xb_j^{*})^{-1} -\Ib\right|^{N_r}} \nonumber \\
&& \hspace{-12mm} - \sum_{i=1}^{M} \frac{P_i}{1-P_0} \frac{1}{\left|
\Ib+\Xb_i\Xb_i^{*}\right|^{N_r}\left|\Ib+\Xb_i\Xb_i^{*}\right|^{-N_r}} \ , \label{STORMpearsonsfinal}
\end{eqnarray}
if $\Ib - \Xb_i\Xb_i^{*}\Xb_j\Xb_j^{*}$ is positive definite $\ \forall \ i,j$, and $\infty$ otherwise. Simplification of (\ref{STORMpearsonsfinal}) results in  $\Delta\left(\overline{P}_{Y}||P_{Y | X =\Zrb}\right)$ given in Theorem \ref{STORMproppearson}.
\endproof
 \end{appendix}

  
\begin{figure}
  \centerline{\scalebox{0.65}{\includegraphics{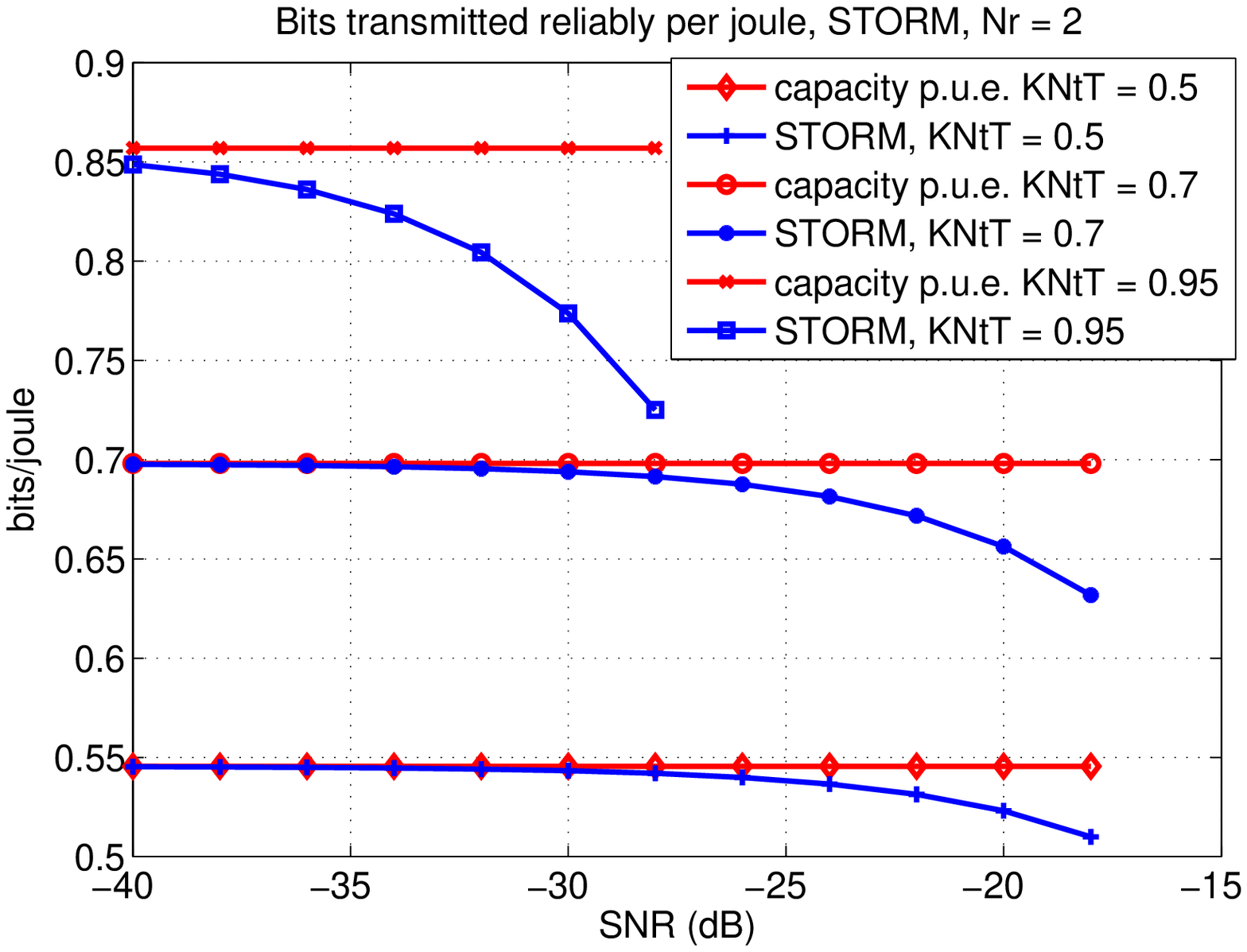}}}
    \caption{Plot of first order approximation of $\frac{I(P)}{P}$ in (\ref{STORMTaylor}) vs. $P$ for STORM, and its convergence to the capacity per unit energy,  for different values of $K N_t T$. } \label{convergence_withP} 
  \end{figure}

  \begin{figure}
  \centerline{\scalebox{0.65}{\includegraphics{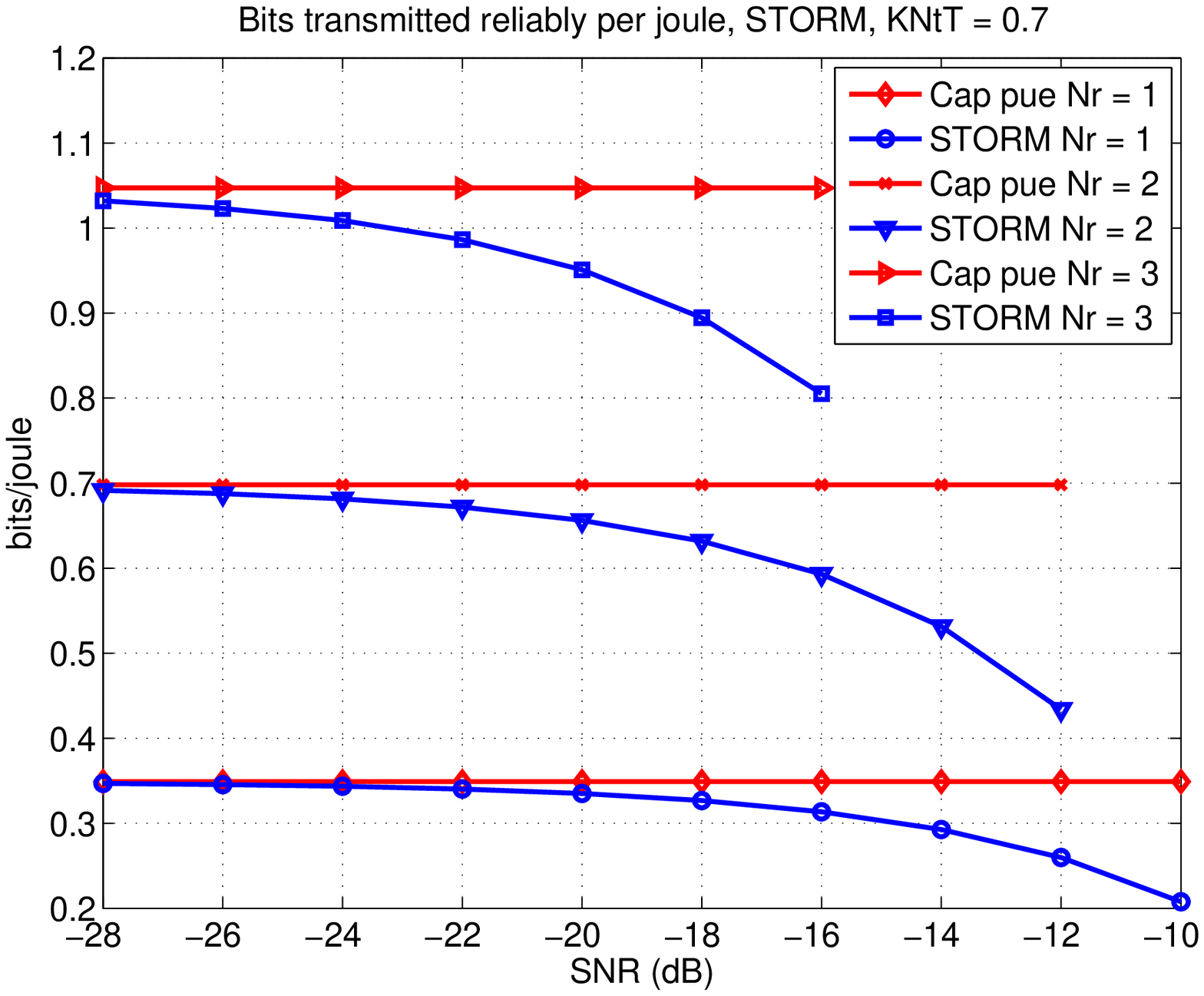}}}
    \caption{Plot of first order approximation of $\frac{I(P)}{P}$ in (\ref{STORMTaylor}) vs. $P$ for STORM,  for different values of $N_r$. }  \label{STORM_Nr} 
  \end{figure}



\newpage 

\noindent {{\large\bf Biography of Shivratna G. Srinivasan}}

Shivratna Giri Srinivasan (S'07) received the B.Tech. degree in electrical engineering
from the Indian Institute of Technology, Madras, in 2002. He received the M.S. and Ph.D. degrees in electrical and computer engineering from the University of Colorado at Boulder in 2005 and 2007, respectively. In September 2007, he joined Qualcomm, Inc., San Diego, as a Senior Engineer and is working on 4G wireless modem designs.

  \vspace{1cm}
 \noindent {{\large\bf Biography of Mahesh K. Varanasi}}

 Mahesh K. Varanasi (S'87--M'89--SM'95) received the Ph.D. degree in electrical engineering from Rice University, Houston, TX, in 1989.  He joined the Electrical and Computer Engineering of the University of Colorado at Boulder in 1989 as an Assistant Professor where he was later an Associate Professor during 1996-2001 and is now a Professor since 2001. His research and teaching interests are in the areas of communication and information theory, wireless communication and coding, detection and estimation theory, and signal processing. He has published on a variety of topics in these fields and is a Highly Cited Researcher in the ``Computer Science'' category according to the ISI Web of Science. He is currently serving as an Editor for the IEEE Transactions on Wireless Communications.


\begin{thebibliography}{10}

\bibitem{ShannonCE:comm}
C.~E. Shannon,
\newblock ``{Communication in the Presence of Noise},''
\newblock {\em Proceedings of the IRE}, vol. 37, no. 1, pp. 10--21, Jan. 1949.

\bibitem{KennedyRS:Fading}
R~S Kennedy,
\newblock {\em Fading Dispersive Communication Channels},
\newblock Wiley and Sons, 1969.

\bibitem{JacobsI:incoherent}
I.~Jacobs,
\newblock ``{The asymptotic behavior of incoherent M-ary communication
  systems},''
\newblock {\em Proceedings of the IEEE}, vol. 51, no. 1, pp. 251--252, Jan.
  1963.

\bibitem{GallagerRG:Info}
R~G Gallager,
\newblock {\em Information Theory and Reliable Communication},
\newblock Wiley and Sons, 1968,
\newblock Section 8.6.

\bibitem{TelatarIE:wideband}
I.~E. Telatar and D.~N.~C. Tse,
\newblock ``{Capacity and mutual information of wideband multipath fading
  channels},''
\newblock {\em IEEE Trans. Inform. Theory}, vol. 46, no. 4, pp. 1384--1400,
  July 2000.

\bibitem{VerduS:efficiency}
Sergio Verd\'u,
\newblock ``Spectral efficiency in the wideband regime,''
\newblock {\em IEEE Trans. Inform. Theory}, vol. 48, no. 6, pp. 1319--1343,
  June 2002,
\newblock Special Issue on Shannon Theory: Perspective, Trends, and
  Applications.

\bibitem{MedardM:bandwidth}
M.~Medard and R.~G. Gallager,
\newblock ``{Bandwidth Scaling for fading multipath channels},''
\newblock {\em IEEE Trans. Inform. Theory}, vol. 48, no. 4, pp. 840--852, Apr.
  2002.

\bibitem{SubraVG:broadband}
V.~G. Subramanian and B.~Hajek,
\newblock ``{Broad-band fading channels: signal burstiness and capacity},''
\newblock {\em IEEE Trans. Inform. Theory}, vol. 48, no. 4, pp. 809--827, Apr.
  2002.

\bibitem{RaoC:lowSNR}
Chaitanya Rao and Babak Hassibi,
\newblock ``Analysis of multiple-antenna wireless links at {Low} {SNR},''
\newblock {\em IEEE Trans. Inform. Theory}, vol. 50, no. 9, pp. 2123 -- 2130,
  Sept. 2004.

\bibitem{PrelovV:secondorder}
V~Prelov and Sergio Verd\'u,
\newblock ``Second-order asymptotics of mutual information,''
\newblock {\em IEEE Trans. Inform. Theory}, vol. 50, no. 8, pp. 1567--1580,
  Aug. 2004.

\bibitem{HajekB:smallpeak}
Bruce Hajek and V~Subramaniam,
\newblock ``Capacity and reliability function for small peak signal
  constraints,''
\newblock {\em IEEE Trans. Inform. Theory}, vol. 48, no. 4, pp. 828--839, Apr.
  2002.

\bibitem{GursoyMC:nonc}
Mustafa~Cenk Gursoy, H.~Vincent Poor, and Sergio Verd\'u,
\newblock ``Noncoherent {R}ician {F}ading channel - {P}art {I}{I}: {S}pectral
  {E}fficiency in the {L}ow-{P}ower {R}egime,''
\newblock {\em IEEE Trans. Wireless Commun.}, vol. 4, no. 5, pp. 2207--2221,
  Sept. 2005.

\bibitem{Abou-FaycalIC:capacity}
Ibrahim~C. Abou-Faycal, Mitchell~D Trott, and Shlomo {Shamai (Shitz)},
\newblock ``The {C}apacity of discrete-time memoryless {R}ayleigh fading
  channels,''
\newblock {\em IEEE Trans. Inform. Theory}, vol. 47, no. 4, pp. 1290--1301, May
  2001.

\bibitem{GursoyMC:capacity}
Mustafa~Cenk Gursoy, Vincent Poor, and Sergio Verdu,
\newblock ``{The Noncoherent Rician Fading Channel - Part I : Structure of the
  Capacity-Achieving Input},''
\newblock {\em IEEE Trans. Wireless Commun.}, vol. 4, no. 5, pp. 2193--2206,
  Sept. 2005.

\bibitem{HuangJ:optdist}
Jianyi Huang and Sean~P. Meyn,
\newblock ``{Characterization and Computation of Optimal Distributions for
  Channel Coding},''
\newblock {\em IEEE Trans. Inform. Theory}, vol. 51, no. 7, pp. 2336--2351,
  July 2005.

\bibitem{SrinivSG:const}
Shivratna~Giri Srinivasan and Mahesh~K Varanasi,
\newblock ``{Constellation Design for the Noncoherent MIMO Rayleigh-Fading
  Channel at General SNR},''
\newblock {\em IEEE Trans. Inform. Theory}, vol. 53, no. 4, pp. 1572--1584,
  Apr. 2006.

\bibitem{BorranMJ:NoncSTC}
M~J Borran, Ashutosh Sabharwal, and B~Aazhang,
\newblock ``On design criteria and construction of noncoherent space-time
  constellations,''
\newblock {\em IEEE Trans. Inform. Theory}, vol. 49, no. 10, pp. 2332--2351,
  Oct. 2003.

\bibitem{MarzettaTL:cap}
T~L Marzetta and Bertrand~M Hochwald,
\newblock ``Capacity of a mobile multiple-antenna communication link in
  {Rayleigh} flat fading,''
\newblock {\em IEEE Trans. Inform. Theory}, vol. 45, no. 1, pp. 139--157, Jan.
  1999.

\bibitem{WozenJM:moddemod}
J.~M. Wozencraft and R.~S. Kennedy,
\newblock ``Modulation and demodulation for probabilistic coding,''
\newblock {\em IEEE Trans. Inform. Theory}, vol. 12, pp. 291--297, July 1966.

\bibitem{MasseyJL}
J.~L. Massey,
\newblock ``{Coding and modulation in digital communication},''
\newblock in {\em Proc. Intl. Zurich Seminar on Communications}, Zurich,
  Switzerland, 1974.

\bibitem{HeroAO:sig}
Alfred~O {Hero, {III}} and Thomas~L Marzetta,
\newblock ``Cut-off rate and signal design for the {Rayleigh} fading
  space--time channel,''
\newblock {\em IEEE Trans. Inform. Theory}, vol. 47, no. 6, pp. 2400--2416,
  Sept. 2001.

\bibitem{AgrawalD:grassman}
D.~Agrawal, T.~J. Richardson, and R.~L. Urbanke,
\newblock ``{Multiple-antenna signal constellations for fading channels},''
\newblock {\em IEEE Trans. Inform. Theory}, vol. 47, no. 6, pp. 2618--2626,
  Sept. 2001.

\bibitem{HochwaldB:sptimesig}
B~M Hochwald, Thomas~L Marzetta, T~J Richardson, W~Sweldens, and R~Urbanke,
\newblock ``Systematic design of unitary space--time constellations,''
\newblock {\em IEEE Trans. Inform. Theory}, vol. 46, no. 6, pp. 1962--1973,
  Sept. 2000.

\bibitem{TarokhV:sptime}
Vahid Tarokh, Nambirajan Seshadri, and A~Robert Calderbank,
\newblock ``Space--time codes for high data rate wireless communications:
  Performance criterion and code construction,''
\newblock {\em IEEE Trans. Inform. Theory}, vol. 44, no. 2, pp. 744--765, Mar.
  1998.

\bibitem{SGSrinivasan:STORMaller}
Shivratna~Giri Srinivasan and Mahesh~K Varanasi,
\newblock ``{STORM}: Optimal constellations for noncoherent {MIMO}
  communications at low {SNR} under {PAPR} constraints,''
\newblock in {\em Proc. Allerton Conf. on Comm. Control, and Comput.},
  Monticello, Illinois, Sept. 2006.

\bibitem{SGSrinivasan:STORMisit}
Shivratna~Giri Srinivasan and Mahesh~K Varanasi,
\newblock ``Mutual information optimal constellations for the low {SNR}
  noncoherent {MIMO} {R}ayleigh fading channel,''
\newblock in {\em Proc. IEEE Intl. Symposium on Information Theory}, Nice,
  France, June 2007.

\bibitem{SethuramanV:MIMO}
Vignesh Sethuraman, Ligong Wang, Bruce Hajek, and Amos Lapidoth,
\newblock ``{Low SNR Capacity of Fading Channels - MIMO and Delay Spread},''
\newblock in {\em Proc. IEEE Intl. Symposium on Information Theory}, Nice,
  France, June 2007.

\bibitem{HorstR:glo}
Reiner Horst, Panos~M. Pardalos, and Nguyen~V. Thoai,
\newblock {\em Introduction to Global Optimization},
\newblock Kluwer, 2000.

\bibitem{BoydS:Optim}
S~Boyd and L~Vandenberghe,
\newblock {\em Convex Optimization},
\newblock Cambridge University Press, Cambridge, U.K., 2004.

\bibitem{VerduS:unitcost}
Sergio Verd\'u,
\newblock ``On channel capacity per unit cost,''
\newblock {\em IEEE Trans. Inform. Theory}, vol. 36, no. 5, pp. 1019--1030,
  Sept. 1990.

\bibitem{WuX:MIMOexp}
X.~Wu and R.~Srikant,
\newblock ``{MIMO Channels in the Low SNR Regime: Communication Rate, Error
  Exponent and Signal Peakiness},''
\newblock {\em IEEE Trans. Inform. Theory}, Apr. 2007.

\bibitem{MarshallAW:ineq}
A.~W. Marshall and I.~Olkin,
\newblock {\em Inequalities: Theory of Majorization and Its Applications},
\newblock Academic Press, New York, 1979.

\bibitem{SethuramanV:cuelowsnr}
Vignesh Sethuraman and Bruce Hajek,
\newblock ``Capacity per unit energy of fading channels with a peak
  constraint,''
\newblock {\em IEEE Trans. Inform. Theory}, vol. 51, no. 9, pp. 3102--3120,
  Sept. 2005.

\bibitem{ZhengL:chcoh}
Lizhong Zheng, David N.~C. Tse, and Muriel Medard,
\newblock ``Channel coherence in the low {SNR} regime,''
\newblock {\em IEEE Trans. Inform. Theory}, vol. 53, no. 3, pp. 976--997, Mar.
  2007.

\bibitem{Rays:noncWB}
Siddharth Ray, Muriel Medard, and Lizhong Zheng,
\newblock ``On non-coherent {MIMO} channels in the wideband regime: Capacity
  and reliability,''
\newblock {\em IEEE Trans. Inform. Theory}, vol. 53, no. 6, pp. 1983--2009,
  June 2007.

\bibitem{RaghavanV:sparse}
Vasanthan Raghavan, Gautham Hariharan, and Akbar Sayeed,
\newblock ``Capacity of sparse multipath channels in the ultra-wideband
  regime,''
\newblock {\em IEEE Journal on Selected Topics in Signal Processing}, vol. 1,
  no. 3, pp. 357--371, Oct. 2007.

\bibitem{HariharanG:ITA2007}
Gautham Hariharan and Akbar Sayeed,
\newblock ``Non-coherent capacity and reliability of sparse multipath channels
  in the wideband regime,''
\newblock in {\em Information Theory and Applications Workshop}, San Diego,
  Jan. 2007.

\bibitem{TurinGL:quad}
G~L Turin,
\newblock ``The characteristic function of hermitian quadratic forms in complex
  normal variables,''
\newblock {\em Biometrika}, vol. 47, pp. 199--201, June 1960.

\end{thebibliography}
  \end{document}